\documentclass[A4]{aa}
\usepackage{amsmath}  
\usepackage{graphicx}
\usepackage{natbib} 
\usepackage{longtable}
\usepackage{lscape}
\bibpunct{(}{)}{;}{a}{}{,} 
\usepackage{txfonts}
\begin{document}
\relpenalty=10000
\binoppenalty=10000
\newcommand{\MPIA}{1} 
\newcommand{\ARI}{2} 
\newcommand{\IfA}{3} 
\newcommand{\CfA}{5} 
\newcommand{\NCU}{6} 
\newcommand{\JHU}{7} 
\newcommand{\PU}{8} 
   \title{Towards a complete stellar mass function of the Hyades}
   \titlerunning{Towards a complete stellar mass-function of the Hyades. I}
   \subtitle{I. Pan-STARRS1 optical observations of the low-mass  {stellar} content}  

  \author{	
       B. Goldman\inst{\MPIA}
          \and
       S.~R\"{o}ser\inst{\ARI}
          \and
        E. Schilbach\inst{\ARI}
          \and
       E.~A. Magnier\inst{\IfA}
          \and
      C. Olczak\inst{\ARI,\MPIA,4}
          \and
      T. Henning\inst{\MPIA}
          \and
       M. Juri\'{c}\inst{\CfA}
          \and \\
       E. Schlafly\inst{\MPIA}
          \and
       W.~P. Chen\inst{\NCU}
          \and
       I. Platais\inst{\JHU} 
          \and
       W. Burgett\inst{\IfA}
          \and
       K. Hodapp\inst{\IfA}
          \and
       J. Heasley\inst{\IfA}
          \and
       R.~P. Kudritzki\inst{\IfA}
          \and \\
       J.~S. Morgan\inst{\IfA}
          \and
       P.~A. Price\inst{\PU} 
          \and
       J.~L. Tonry\inst{\IfA}
          \and 
       R. Wainscoat\inst{\IfA}
  }

   \offprints{B.~Goldman, {\tt go{\,\hspace{-1pt}}ld\hspace{-1pt}\,man{\,}@mp{}ia.de}}

   \institute{
              Max Planck Institute for Astronomy, K\"onigstuhl~17, D--69117 Heidelberg, Germany
         \and 
              Astronomisches Rechen-Institut, Zentrum f{\"u}r Astronomie der Universit{\"a}t Heidelberg, M{\"o}chhofstrasse 12--14, D--69120 Heidelberg, Germany
         \and 
           Institute for Astronomy, University of Hawaii, Honolulu, HI 96822, U.S.A. 
         \and 
           National Astronomical Observatories of China, Chinese Academy of Sciences (NAOC/CAS), 20A Datun Lu, Chaoyang District, Beijing 100012, PR China
         \and 
           Harvard-Smithsonian Center for Astrophysics, 60 Garden Street, Cambridge, MA 02138, U.S.A 
         \and 
           Graduate Institute of Astronomy, National Central University, Jhongli 32054, Taiwan
          \and 
           Department of Physics \& Astronomy, The Johns Hopkins University, Baltimore, MD 21218, U.S.A.
          \and 
           Department of Astrophysical Sciences, Princeton University, Princeton, NJ 08544, USA
       }

   \date{Received / Accepted}

   \abstract
   {The Hyades cluster is an ideal target to study the dynamical evolution of a star cluster over the entire mass range due to its intermediate age and proximity to the Sun. }
   {We wanted to extend the Hyades mass function towards lower masses down to $0.1\,{\rm M_{\odot}}$ and to use the full three-dimensional spatial information to characterize the dynamical evolution of the cluster.} 
   {We performed a kinematic and photometric selection using the PPMXL and Pan-STARRS1 sky surveys, to search for cluster members up to 30 pc from the cluster centre. We determined our detection efficiency and field star contamination rate to derive the cluster luminosity and mass functions down to masses of 0.1\,M$_\odot$. The thorough astrometric and photometric constraints minimized the contamination.
   A minimum spanning tree algorithm was used to quantify the mass segregation.
  }
   {{We discovered {43}~new Hyades member candidates with velocity perpendicular to the Hyades motion up to 2\,$\rm km.s^{-1}$. They have mass estimates between {0.43} and 0.09\,M$_\odot$, for a total mass of {10}\,M$_\odot$. This doubles the number of Hyades candidates with masses smaller than 0.15\,M$_\odot$. We provide an additional list of {11}\,possible candidates with velocity perpendicular to the Hyades motion up to 4\,$\rm km\,s^{-1}$. } The cluster is significantly mass segregated. The extension of the mass function towards lower masses provided an even clearer signature than estimated {in the past}. We also identified as likely Hyades member an L0 dwarf previously assumed to be a field dwarf. {Finally we question the membership of a number of previously published candidates, including a L2.5-type dwarf.} }
  {}
  
   \keywords{Open clusters and associations: individual: Hyades -- Stars: luminosity function, mass function -- Stars: low mass, brown dwarfs -- Stars: individual: 2MASSI~J0230155+270406, 2MASS J05233822-1403022
   }
  
   \maketitle
%

\section{Introduction}

The determination of the (present day) mass functions of stellar clusters shed light on the formation of stars and brown dwarfs and of the production of the field population out of the stellar nurseries.
The evolution of the cluster mass function, estimated from the study of various clusters of different ages, may be interpreted as the consequence of dynamical interactions of the cluster members and of the cluster with the Galactic gravitational potential.
Studies such as \citet{Bouvi08,Wang11,Boudr12} of the intermediate-age clusters Pleiades, Praesepe and Hyades, have revealed the role of evaporation of low-mass members, as well as possibly different stages in the evolution at a given age.
 
Because of its proximity to the Sun  \citep[about $47$\,pc, for the most recent determinations see, e.g. ][] {vLeeu09,McArt11} and intermediate age \citep[about $650$\,Myr, see e.g.  ][]{Perry98,DeGen09}, the Hyades cluster is a valuable target to study the evolution of stellar clusters.

It is desirable to extend the Hyades census to lower masses to provide new benchmark objects at the end of the main sequence, with known age and metallicity. They can further constrain evolutionary models of the stellar core, of the stellar rotation \citep{Delor11},  {and other stellar properties}.

{The Hyades cluster has a relatively large space velocity of $(U,V,W)=(-41,-19,-1)\rm\,km\,s^{-1}$. This allows us to efficiently remove contaminants based on the proper motion. }
\citet[][R11 thereafter]{Roese11} searched the vicinity of the Hyades cluster up to 30\,pc from the centre, using the convergent point method and the wide-area surveys PPMXL \citep{Roese10} and CMC14 \citep{Copen06}.

{The kinematic selection needs to be complemented by a photometric selection in order to improve the purity of the candidate sample. } {The \underline{Pan}oramic \underline{S}urvey \underline{T}elescope \underline{A}nd \underline{R}apid \underline{R}esponse \underline{S}ystem (Pan-STARRS) will eventually comprise multiple telescopes and cameras, and currently consists of a single telescope and camera known as Pan-STARRS1 \citep[or PS1, ][]{Kaise02}.}
PS1\footnote{\tt www.ps1sc.org} is a new optical/red survey instrument with a large figure of merit thanks to its 7-deg$^2$ field of view, that provides new multi-epoch, multi-band observations of  all the sky north of $-30\degr$. 
It uses a dedicated 1.8-m telescope located in Haleakal\={a} on Maui. The pixel scale is 0.25\,arcsec.
It offers the perfect data set to study nearby clusters and extend their membership census to large cluster radii. PS1 has accumulated {more than two years} of survey data, with multiple $grizy$ coverage of the Hyades. 
This data set allows us to push the study of R11 to greater depths and lower masses, and refine the member selection using multi-band photometry and astrometry unavailable until now.


The membership selection is a three step process. First, we select kinematic members by the convergent point method. Second, we further restrict the sample of candidates by photometric selection. Finally, we verify the PPMXL proper motions for all the candidates that survived the kinematic and photometric selections.

The paper is organized in the following way: 
we describe our data set in Sect.\,\ref{obs}.
Section\,\ref{kinsel} presents our kinematic selection while Sect.\,\ref {csel} presents our photometric  selection.
Candidates of particular interested are discussed in Sect.\,\ref{inters}.
The corrections to derive the luminosity function are described in Sect.\,\ref{lf}, whereas the mass function is presented in Sect.\,\ref{mf}. 
Finally we study the spatial distribution of our candidates in Sect.\,\ref{spatstruc}.

Through-out the article we conform to the usual practice and report optical photometry in the AB system (PS1, SDSS surveys) and near- and mid-infrared  {photometry} in the Vega system (2MASS, UKIDSS, WISE surveys).


\section{Observations} \label{obs}

\subsection{Pan-STARRS1 observations of the Hyades} \label{PS1obs}

A fraction of 56\% of the observing time of PS1 is dedicated to the so-called $3\pi$ survey, which monitors the sky north of $-30\degr$. 
This survey covers the area in five filters, $g_{\rm P1}, r_{\rm P1}, i_{\rm P1}, z_{\rm P1}$ and $y_{\rm P1}$, with repeated observations over 3\,years. 
{Details about the $3\pi$ survey are in Chambers et al. (in {\it prep.}).} 
The profiles of the blue filters, $g_{\rm P1}, r_{\rm P1}, i_{\rm P1}$, are close to those of the SDSS filters \citep[][]{Stubb10,Tonry12}.
In each filter, two visits of two exposures each are scheduled every year.
The two exposures 
are separated by 30--90\,min, in order to detect fast-moving objects such as near-Earth objects. 
The telescope is pointed to the same direction within a few arcseconds, while the two visits are separated by a few days in the optical ($gri$), up to several months in the red bands ($zy$) to improve the parallax factor range \citep{Beaum10}.
The static fill factor of the camera, taking into accounts the gaps between and within the chips, as well as areas of poor sensitivity, is about 75\%.
As the survey progresses, the camera pointing and orientation is shifted to cover the areas missed in the previous seasons, and optimally improve the photometric and astrometric calibrations.

{Between February 1st, 2010 and August 31, 2012, which form the photometric data we use in this article}, Pan-STARRS1 has observed most of the area of the $3\pi$ survey. 
{With two observing seasons}, most of the Hyades cluster (within $45\degr$ of the centre) has been observed {4 to 6} times in each filter {under photometric or nearly-photometric conditions}, with typical depths of 22\,mag (AB, 50\% completeness limit, for {individual exposures in} $gri$, see section \ref{PSdepth}).
Extended periods of poor weather between mid-October and mid-January have prevented us from completing the program in some limited areas (see Sect.\,\ref{Sec:skycov}).
The image quality varies from FWHM=1.4\arcsec\ in the $g$ band to 1.2\arcsec\ in the $z$ band, averaged over the whole Hyades area.

{The images are processed on a nightly basis by the Image Processing Pipeline \citep[IPP, ][]{Magni06}. The pipeline detrends and debiases the images, searches for sources $5\sigma$ or more above background, and measures PSF photometry and various parameters.
 We select stars with good quality flags\footnote{Stars with good PSF fit, good background measurement, that are not saturated, not falling on chip defects, cosmic rays, diffraction spikes, electronic ghosts, or gaps.}.
}

The data are photometrically calibrated according to the algorithm of \citet{Schla12}, who use repeated observations of the same stars to constrain a model for the PS1 system throughput.  
The technique is an extension of that of \citet{Padma08}, used to calibrate the SDSS.  
Internal comparisons and comparison with the SDSS indicate that the photometric calibration is accurate at the $<1$\%\ level.

{We transform the SDSS photometry into the PS1 photometric system using linear colour transformations (Finkbeiner, {\it priv.com.}). 
We use the PS1 photometric system for candidate selection and in figures, but provide the original SDSS photometry in tables.}

\subsection{Sky coverage}  \label{Sec:skycov}

Both PS1 and SDSS have incomplete sky coverage of the Hyades, 
the former because it is on-going and because of adverse weather pattern in the {past two winter seasons}; 
the latter because the cluster falls outside of the contiguous SDSS Northern Cap  coverage.

To perform a spatial distribution analysis, it is necessary to create maps of the actual coverage. For PS1, the coverage variations have high spatial frequency, while SDSS is mostly contiguous over isolated stripes.

Because the incomplete sky coverage has its origin in the geometry of the focal plane (PS1) or of the survey (SDSS), we consider that the coverage fraction is independent of the candidate brightness.

\begin{figure}
\includegraphics[width=.5\textwidth]{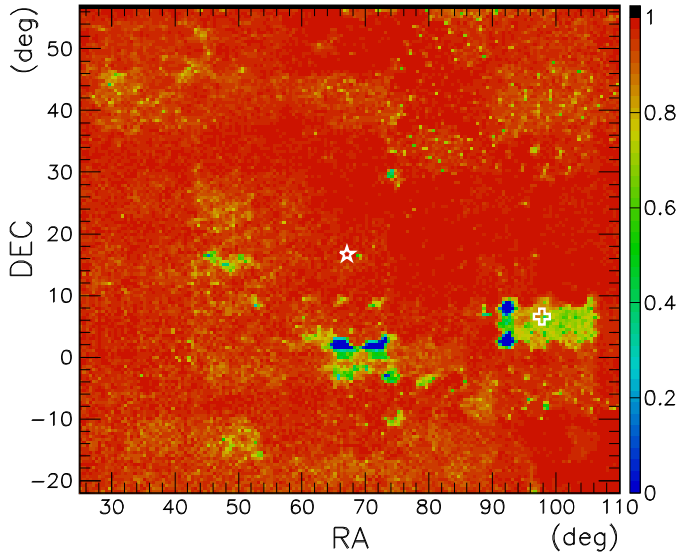}
\caption{Sky coverage of our PS1 data combined with SDSS--{DR8} around the Hyades cluster in either the $g$, $r$ or $i$ filters. The colour scale indicates the fraction of 2MASS stars recovered in our catalogue (see text). The PS1 3-deg-diameter footprint is visible in a few locations. 
The white star indicates the cluster centre and the cross, the convergent point. }
\label{PS1cov}
\end{figure}

About {25}\% of each exposure footprint is not observed because of the gaps between the sensitive devices and the masked areas due to bright stars, trails, etc.
Therefore, a fraction of the kinematic candidates is not detected even if the whole area were covered a few times. 
The PS1 observations are conducted in pairs, called TTI pairs, with an identical pointing within a few arcseconds. The masks of the two images of the pair are thus largely overlapping. 
In our static mask, chip gaps appear every 20\,arcmin, cell gaps every 2.5\,arcmin. To produce a map with sufficient spatial resolution, we search the PS1 catalogue for matches for all 2MASS stars over the area of interest. We then calculate over (12\,arcmin)$^2$ areas the fraction of 2MASS stars recovered in $g$, $r$ and/or $i$ (see Fig.\,\ref{PS1cov}).

This approach could lead to biases and underestimate the sky coverage in regions where 2MASS is deeper than PS1, i.e. in high extinction regions. 
We limit the 2MASS star faintness to $J<15$\,mag to minimize this effect.

This calculation indicates a coverage in any of the $g$, $r$ or $i$ bands of {94}\%, versus only {44}\% in all three bands. For this reason, we chose to select candidates in any band to improve coverage, at the cost of purity. 

We perform a similar analysis for the SDSS and PPMXL surveys. 
The SDSS sky coverage is combined with that of PS1. 
For PPMXL, the average detected fraction of 2MASS stars with magnitudes between 9 and 13 is 98\% (over (30\,arcmin)$^2$ areas).
The lacking areas in PPMXL cover 1.2\,deg$^2$ and are close to very bright stars and M77. 
We do not correct our coverage maps for PPMXL heterogeneities.

\subsection{Depth} \label{PSdepth}

We determine the detection efficiency as a function of magnitudes and sky location of the surveys we use, namely PPMXL and Pan-STARRS1, by comparing them with SDSS DR8.
For each SDSS star ({\tt type=6}), we search within 1\arcsec\ of its position for a PPMXL or a PS1 $g$, $r$ or $i$ detection.
We report the recovery rate as a function of SDSS $g$, $r$ or $i$ magnitudes.
While the SDSS~DR8 catalogue covers only a fraction of the area of interest, this procedure still gives a good estimate of the typical depth of both surveys over the survey region.

This method is not affected by the fact that SDSS is not much deeper than PS1. 
What matters is the purity of the SDSS point-source catalogue, so that spurious SDSS detections are not interpreted as objects missed by PS1.

\begin{figure}
\includegraphics[width=.5\textwidth]{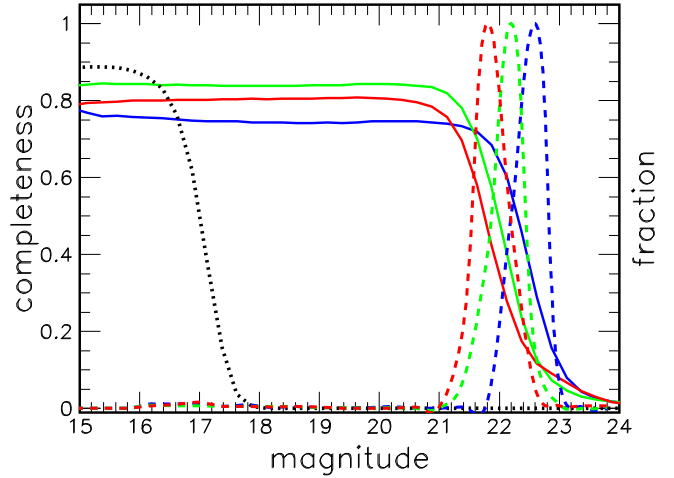}
\caption{Completeness of PS1 as a function of $g$ (blue), $r$ (green), and $i$ (red) SDSS magnitudes (solid curves, right to left), and histograms of the 50\% completeness limit (dashed curves, normalized to unity). The similar completeness of CMC14 as a function of $r$ SDSS magnitude is shown (dotted curve, black).
The maximum reflects the fraction of the area observed in each filter.}
\label{depthPS1}
\end{figure}

We determine the 50\% completeness limit, {(after removing the partial coverage incompleteness),} over $\Delta {\rm RA}\times\Delta {\rm Dec}=1\times 1\,{\rm deg}^2$ areas (see Fig.\,\ref{depthPS1}). 
In the case of PS1, we consider the 50\% completeness relative to the completeness over the 16--18-magnitude range. 
Because of the sharp decline in the PS1 recovery rate, the details of the normalization procedure does not affect much the measured depths.

We find that PS1 has a 50\% completeness limit of \mbox{$g=22.6$}, \mbox{$r=21.2$} and \mbox{$i=21.8$\,mag}, with respective widths of 0.44, 0.48 and 0.33\,mag, over the region of interest (see Fig.\,\ref{depthPS1}).
This is the depth of individual exposures, whose catalogues are combined (unique detections among multiple exposures are considered valid) on a filter per filter basis.

The limiting survey in this study is PPMXL ({with a 50\% completeness limit of $r=20.3$\,mag}, see Sect.\,\ref{PPMXLdepth}), except in some very  {restricted} areas with poor PS1 observing conditions. 
Therefore we ignore the PS1 depth variations over the sky. 

We perform the same analysis for CMC14, {which is used to select low-mass Hyades candidates by}  R11. 
The result, shown in Fig.\,\ref{depthPS1}, is in agreement with the CMC14 documentation.
The CMC14 and PS1 surveys complement each other: the PS1 saturation limit is brighter than CMC14's 95\% completeness limit, and the PS1 depth is about 4\,mag fainter.
In addition, CMC14 does not cover our survey area north of $+50\degr$ and a small area at the edge of the cluster. 

\subsection{Other surveys}

We complement the Pan-STARRS1 photometry and the 2MASS magnitudes from PPMXL {with SDSS  DR8} \citep{SDSS_DR8} and UKIDSS DR8 \citep{Lawre07,Hambl08}, {searching within 1\arcsec\ of the predicted PPMXL position}.  
We report the  SDSS 
photometry  (see Table\,\ref{CandParam}) and use the optical data for the photometric selection.

The SDSS--DR8 coverage of the Hyades is limited to a few 2.5-deg-wide stripes, {making up less than {16}\% of the area. One of the stripes overlaps} the centre. We use the SDSS photometry to {complement} the PS1 coverage, and to check the depth 
of the current PS1 catalogue.

UKIDSS, in its Galactic Cluster Survey component, also covers the Hyades cluster. Because our original kinematic selection requires 2MASS detection, UKIDSS data mostly allows us to refine the photometry of the faintest candidates (see Sect.\,\ref{furthersel}).

Finally, WISE released its {all-sky catalogue on March 14, 2012}\citep{Wrigh10}. We {marginally use} the mid-infrared photometry to select our candidates, and provide their photometry in the WISE filters (cf. Table\,\ref{CandParam}).

\section{Kinematic selection} \label{kinsel}

\subsection{The Convergent Point Method}

The convergent point method \citep[CPM; for a detailed recent description, see][]{vLeeu09} is an important tool to isolate (from the field) an ensemble of stars sharing the same 3D-space motion such as the stars in an open cluster. R11 have applied this method, using a subset of PPMXL having CMC14 $r$-band photometry, to find an empirical main sequence of present-day (and also of formerly bound) members of the Hyades down to an absolute magnitude of $M_{K_S}$ of about 9 mag.
We follow the approach of R11 (Sect.\,4.1) and select only stars within a radius of 30\,pc around the centre of the cluster, {which we define at RA=4h\,28\arcmin 25.7\arcsec, Dec=+16$^{\rm o}$\,42\arcmin 45\arcsec} {(J2000.0)}. 

{We allow a limit of $4\,\rm{km\,s}^{-1}$ for the velocity component $|\rm{v}_\perp|$ perpendicular to the direction to the convergent point {to select possible members}. 
We also set an upper bound to the angle between the proper motion vector and the vector in the direction to the convergent point, of $9.5\degr$,  in order to limit the contamination for candidates closer to the convergent point. }
{In order to increase the purity of our sample, we further require our candidates to have $|\rm{v}_\perp|<2\,\rm km\,s^{-1}$.
We find that these parameters offer a good compromise of purity and completeness:
The stronger constraint roughly decreases by a factor of two the field contamination {(based on our Besan\c{c}on simulation, see Section\,\ref{model})}.
On the other hand, 
a residual velocity of 2\,$\rm km\,s^{-1}$ is still much larger than the velocity dispersion in the cluster, which is predicted to be smaller than 1\,$\rm km\,s^{-1}$ (see R11; {Ernst, \it priv. com.}).

However, the precision on our measurement of $|\rm{v}_\perp|$ itself varies between 1 and 2\,$\rm km\,s^{-1}$. 
{We estimate the $|\rm{v}_\perp|$ error by propagating the PPMXL proper motion measurement errors, and therefore it} depends on target brightness and number of available epochs.
For the faintest candidates close to the PPMXL detection limit, the incompleteness would be significant. 
For instance, the 15\% of kinematic candidates with the largest errors ($\sigma(|\rm{v}_\perp|)>1.5\rm\,km\,s^{-1}$)  have mostly a mass smaller than $0.1\,\rm{M}_\odot$, and are located 10\,pc or more {behind} the cluster centre. 
The incompleteness for such members would raise to 30\%, but as we argue below (see Sec.\,\ref{PPMXLdepth}), the present work concentrates on the mass range above this limit, for which the incompleteness is generally smaller than 10\%.
{Furthermore, Hyads in multiple systems have an orbital motion added to their secular motions.
For wide binaries of the appropriate period, in the order of the PPMXL time baseline (a few decades) or longer, the proper motions of the components will differ from the proper motion of the centre of gravity, and the perpendicular velocity of each component may be larger than 2\,km/s \citep[see footnote\,\ref{77Tau} in Sect.\,\ref{spatstruc} for an example; also][]{Wiele99,Frank07}.
}

We provide the list of candidates with $|\rm{v}_\perp|<2\rm\,km\,s^{-1}$ and those with $2<|\rm{v}_\perp|<4\rm\,km\,s^{-1}$ together in the Tables\,\ref{NewCand} and 6\footnote{Table\,6 is only available in electronic form at the CDS via anonymous ftp to cdsarc.u-strasbg.fr (130.79.128.5) or via http://cdsweb.u-strasbg.fr/cgi-bin/qcat?J/A+A/}.} 

As R11 describe, the CPM is no final confirmation for a star to be co-moving with the bulk of the Hyades cluster, it ''predicts'' a secular parallax and a radial velocity for each candidate. 
Both quantities have to be verified to conclusively confirm membership. 
Using the predicted secular parallaxes one is able to construct colour-absolute magnitude diagrams (depending on the observed bands) to check if the candidates populate allowed loci in these diagrams.

The loci of M-stars in the $M_{K_s}$ vs. $J - K_s$ form an almost vertical line for $M_{K_s}\geq$\,5.5\,mag with the consequence that the predicted $M_{K_s}$ cannot be checked unless observations in a shorter wavelength band is available.
{Optical photometry from PPMXL itself is not useful to check the loci of kinematic candidates in a colour-magnitude diagram. 
It originates from photographic photometry from USNO-B based on POSS I and II, which is not accurate enough for this purpose.} 
Hence, the R11 search for members turned out to be limited by the completeness limit ($r \approx 17$\,mag) of CMC14.

The full PPMXL catalogue goes considerably deeper than CMC14 (Figs.\,\ref{depthPS1} and \ref{depthPPMXL}), so including PS1 enables to check all the low-mass kinematical candidates for photometric membership. In this paper we selected {16543}\ such candidates, with partial overlap with the list published in R11. 
For the latter objects, the deeper PS1 photometry will refine the results based on the shallower CMC14 catalogue.

\subsection{Confirmation of the PPMXL astrometry} \label{confastrom}

The accuracy and the reliability of the proper motions are essential in the CPM. \citet{Roese10} state that PPMXL contains a large number of stars with spuriously large proper motions, many of them are related to the (about 10 \%) double entries in PPMXL. 
We can verify the PPMXL proper motions by the present-day PS1 astrometry itself.
PPMXL uses USNO-B1.0 positions, complemented with 2MASS, with epochs spread over five decades for the objects detected in the first plate surveys, down to two decades for the objects only detected in the latest plate surveys. PS1 adds measurements 12\,years on average after 2MASS's epoch.

\begin{figure}
\includegraphics[width=.5\textwidth]{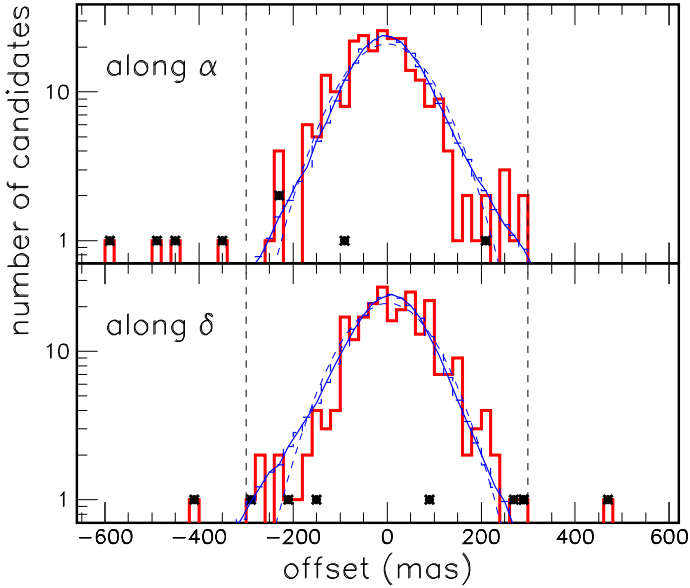}
\caption{Histograms of the offsets along RA and Dec between the expected and observed positions of the {247} candidates (thick line) and nearby field objects (normalized; thin line, with the fitted Gaussian as dashed line). The {eight} candidates with an offset larger than 300\,mas are highlighted. }
\label{astrom}
\end{figure}

The current PS1 astrometric registration is performed on a chip-per-chip basis (20-arcmin on a side) against 2MASS \citep{Magni08}. 
{Then the astrometric solutions are improved with PS1 detections alone, using overlapping PS1 exposures.}
The typical precision against 2MASS is 80\,mas in each direction. 
However here we are interested in the PS1 precision relative to PPMXL (or vice-versa). To estimate it we search the PS1 catalogue (all bands {including $z_{\rm P1}$ and $y_{\rm P1}$)} for PPMXL detections, {up to 12\,arcmin of each candidate}.   
We only consider objects with proper motion smaller than 5\,mas\,yr$^{-1}$ in both directions, to avoid erroneously large PPMXL proper motions due to confusion.
We measure the offset between the expected positions based on PPMXL astrometry (proper motion and positions at its published epoch 2000.0) and the PS1 {observed} positions {on each exposure}. 
{Finally, we subtract from those offsets the mean offset of the objects in the vicinity of each candidate.
This brings the PS1 astrometric system, which is relative to 2MASS and to the referential of the field bright stars {and not to the inertial referential of extragalactic sources}, into the IRCS. (Over 11\,years, the correction is typically $\Delta \alpha=-23\rm\,mas\,yr^{-1}$ and $\Delta \delta=-47\rm\,mas\,yr^{-1}$ with a dispersion of $35\rm\,mas\,yr^{-1}$). }
{We find that the core of the offset distributions along each direction, $\alpha$ and $\delta$, are well fitted by a Gaussian function of width 90\,mas, which is therefore the dispersion of the PS1 astrometry relative to PPMXL (Fig.\,\ref{astrom}, blue curves).}

Similarly, for each kinematic candidate, we extrapolate its PPMXL position using the PPMXL proper motion to the epoch of observations, and search for detections within 1\,arcsec of that position. 
This search radius is much larger than the typical accuracy measured above, so that we do not need to use an adaptive search radius as a function of the PPMXL position or proper motion uncertainties. 
On the other hand, erroneous large proper motions from PPMXL (wrong by more than 100\,mas\,yr$^{-1}$) would prevent us from recovering the (actually slow-moving) candidates.
{Those objects could not be Hyades members. }
 
We find that most objects with an offset larger than {300\,mas are likely slow}-moving, distant sources. 
{A few remaining candidates} with erroneous proper motions due to confusion, {partly} identified by eye examination, can have small offsets if they have proper motions smaller than $60$\,mas\,yr$^{-1}$. 
With these limitations in mind, we require that the offsets be smaller than {300\,mas, 
measured with four or more PS1 measurements.}
 
We do not use other surveys such as SDSS or UKIDSS to perform a similar verification, as the baseline between the 2MASS observations and SDSS or most of UKIDSS imaging is only a few years and do not set strong constraints.

\section{Selection of Hyades member candidates} \label{csel}

The kinematic sample is affected by background and foreground stars with the same {proper motions} mimicking the Hyades space velocities for their sky location. 
To remove these contaminants, we use optical photometry from Pan-STARRS1 and SDSS, {combined with infrared photometry from 2MASS,  and WISE}. 

\subsection{Photometric selection} \label{optselection}

We again follow the steps of R11 and select our candidates in the absolute magnitude vs. colour diagrams, {taking the secular parallax from the CPM.}  

\begin{figure*}
\includegraphics[width=.5\textwidth]{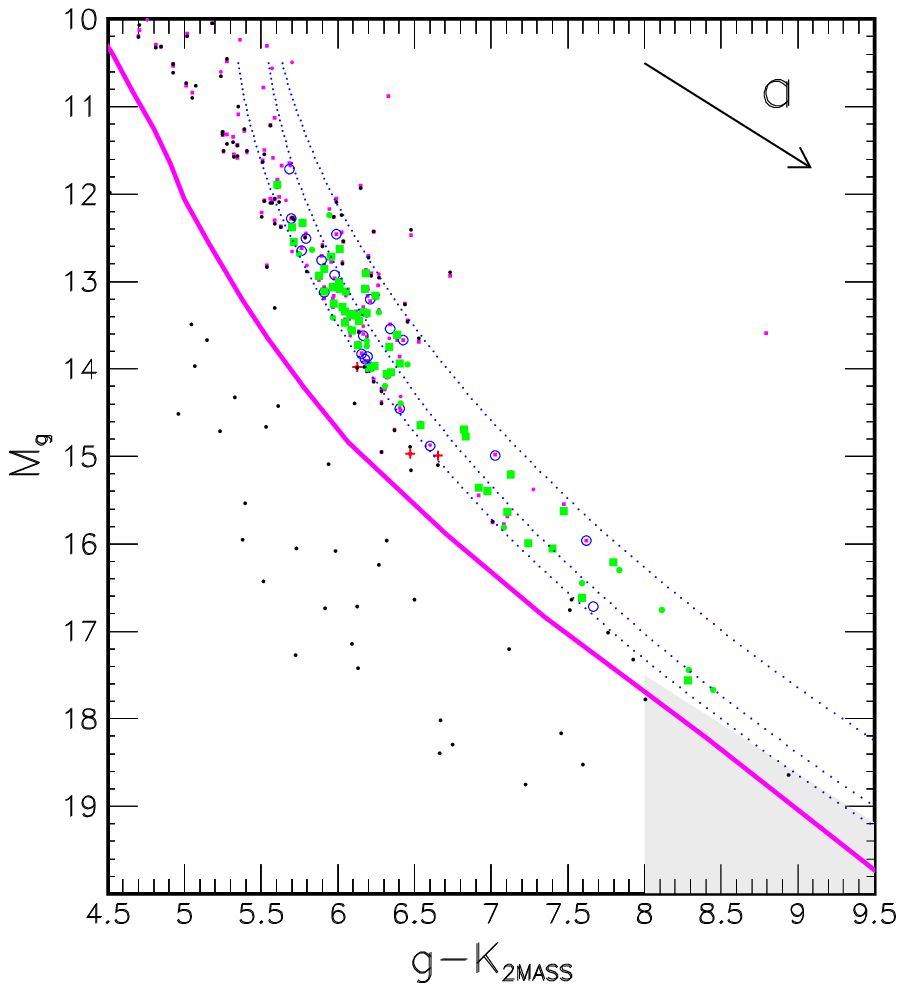}
\includegraphics[width=.5\textwidth]{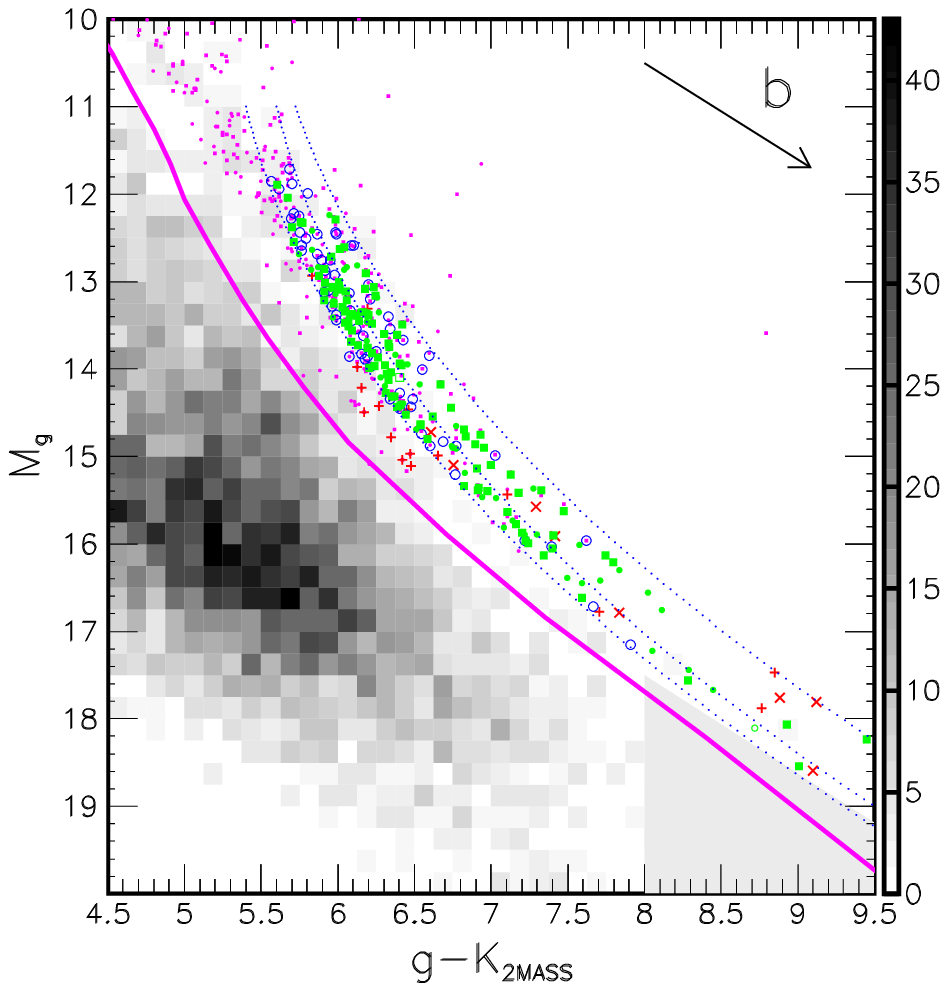}
\includegraphics[width=.5\textwidth]{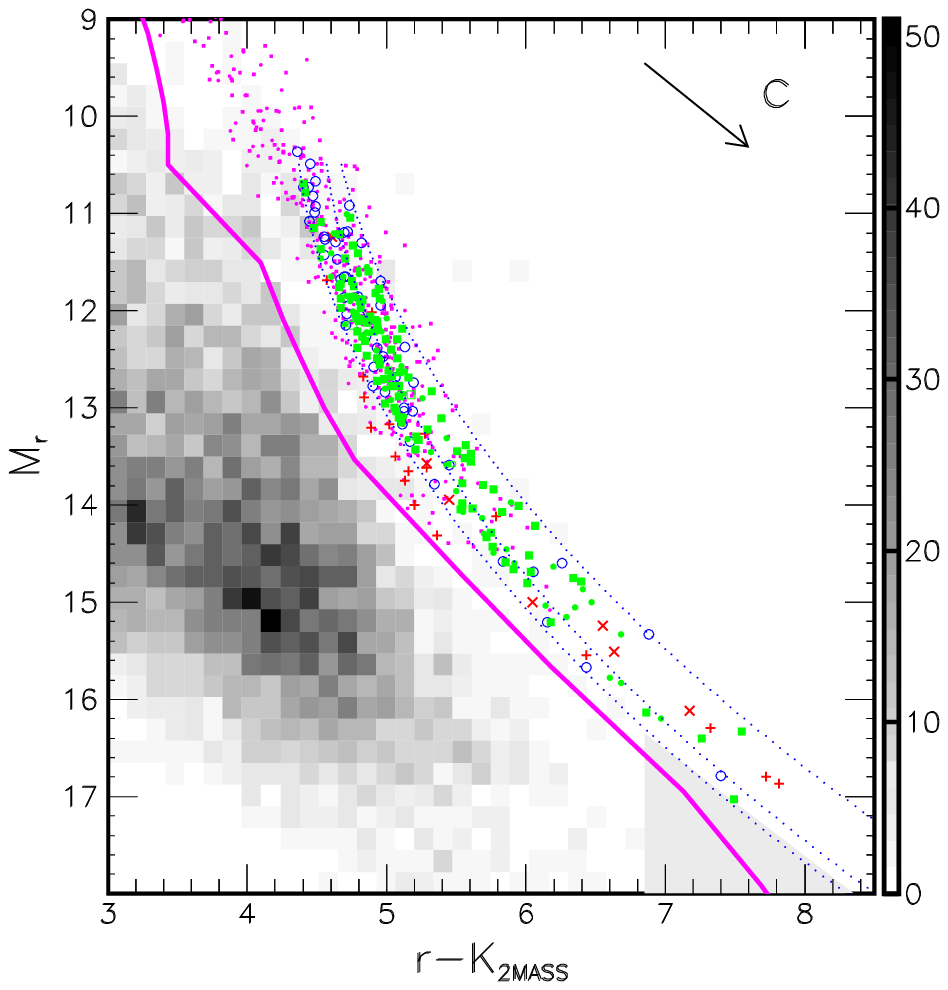}
\includegraphics[width=.5\textwidth]{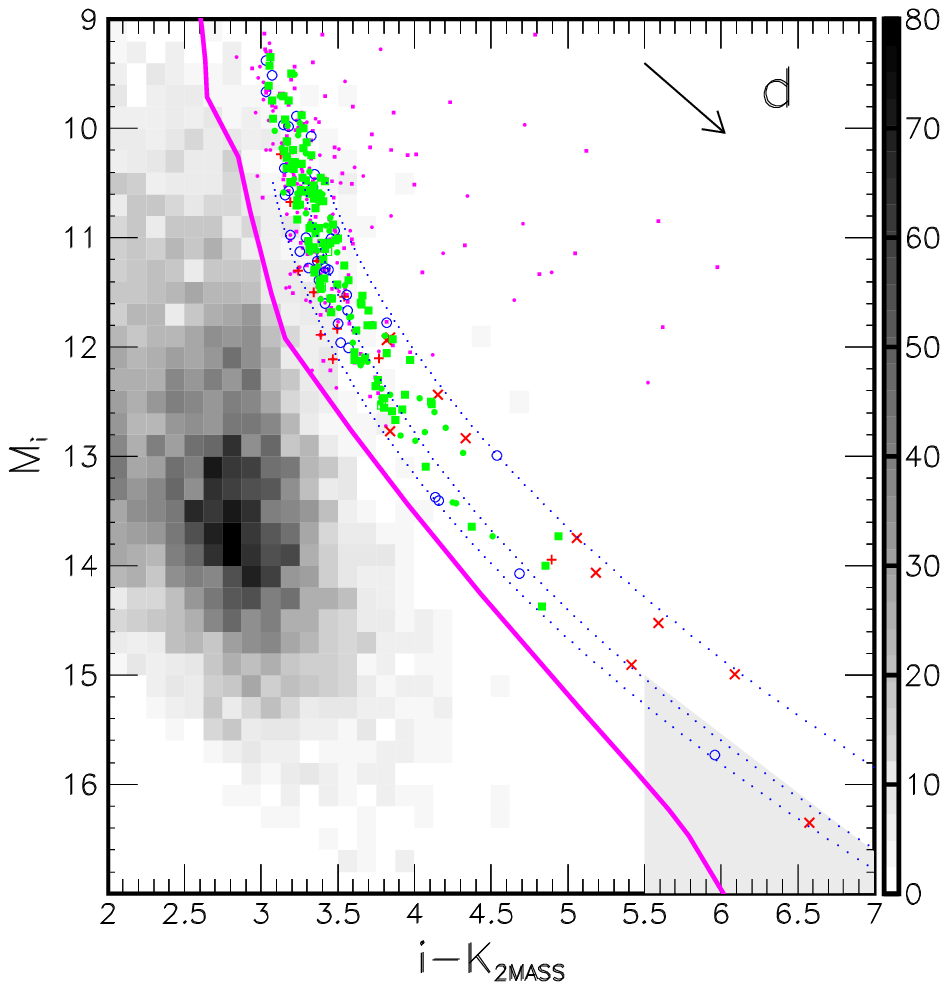}
\caption{
{\it a)} Colour-magnitude diagram of $g_{\rm P1}-K_{\rm 2MASS}$ for all the kinematic candidates with $|v_\bot|<2$\,km\,s$^{-1}$ within 9\,pc of the cluster centre. {\it bcd)} {Hess diagrams} of $g_{\rm P1}-K_{\rm 2MASS}$, of $r_{\rm P1}-K_{\rm 2MASS}$, and of $i_{\rm P1}-K_{\rm 2MASS}$, for the all kinematic candidates, i.e. with $|v_\bot|<4$\,km\,s$^{-1}$ {up to 30\,pc from the cluster centre}. \newline
Filled (green) circles indicate candidates selected in $g$, $r$ and $i$; open (blue) circles: candidates selected in one or two filters; (red) crosses and pluses: candidates suspected to be giants (see Fig.\,\ref{JHHK}) or to have bad proper motion measurements. Dots (magenta) represent R11 Hyades candidates with CMC14 $r$-band photometry in {\it c} and PS1 and SDSS photometry in {\it a, b, d}. 
The dotted lines (blue) indicate our selection zone, while the thick (magenta) line represents the BT-Settl model 600-Myr isochrone. 
The grey area at the faint, red edge shows our incompleteness limit. } 
\label{CMDs}
\end{figure*}

We define our selection box empirically {in colour-magnitude diagrams} (see Fig.\,\ref{CMDs}). 
Indeed, the available theoretical isochrones do not reproduce accurately the colours of the Hyades low-mass stars. 
{As an example, we display the latest calculation (CIFIST2011) of the BT-Settl 600-Myr isochrone (F.~Allard, {\it priv.com}).\footnote{\tt http://phoenix.ens-lyon.fr/Grids/BT-Settl/\newline CIFIST2011/ISOCHRONES/} }
On the other hand, each colour-magnitude diagram clearly exhibits a sequence above the cloud of field stars, {particularly for $M_g<16$\,mag}. 
{In Fig.\,\ref{CMDs}a we show the kinematic candidates with $|v_\bot|<2$\,km\,s$^{-1}$ within 9\,pc of the cluster centre. Close to the cluster centre the field contamination is minimal and the over-density clearly visible.
 Most of the field stars do not share the cluster motion, so that their kinematic parallax is erroneous and their calculated absolute magnitude does not correspond to that of a Hyades-sequence star of their colour. }
In Fig.\,\ref{CMDs} {\it c}, the locus of the R11 candidates (with CMC14 $r$-band photometry) overlaps with this sequence. 

\begin{figure}
\includegraphics[width=.5\textwidth]{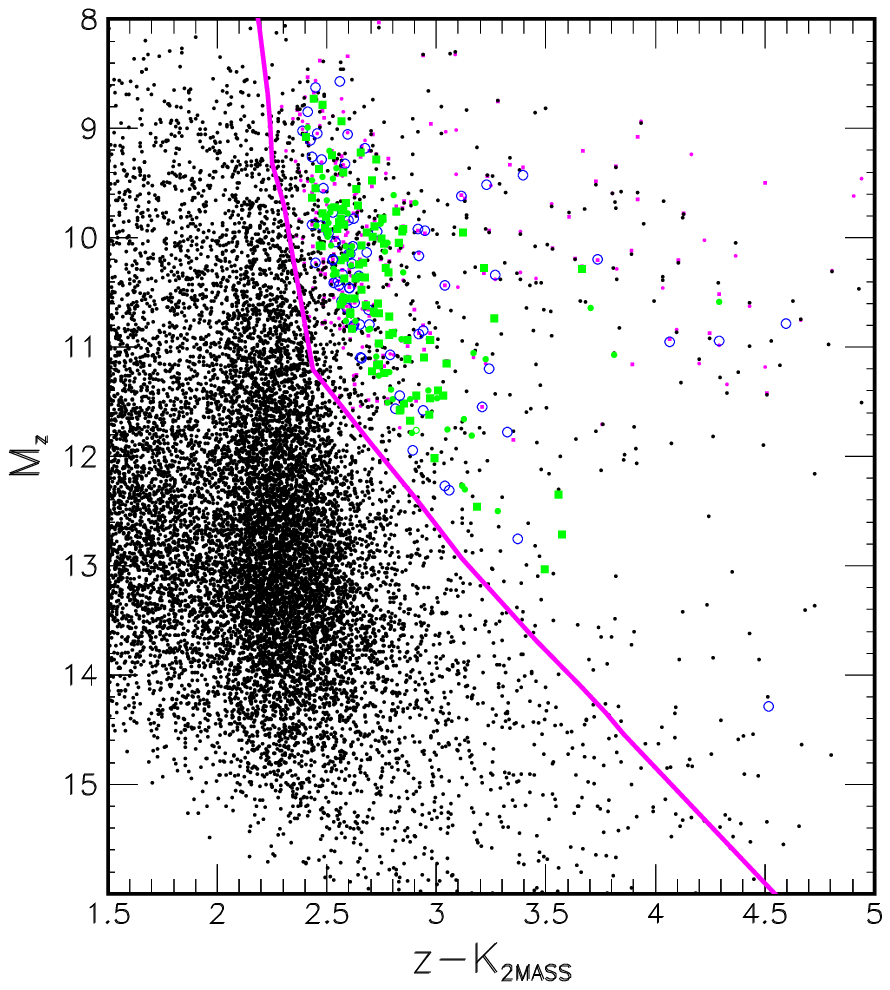}
\caption{
Colour-magnitude diagram of $z_{\rm P1}-K_{\rm 2MASS}$ for all the kinematic candidates, i.e. with $|v_\bot|<4$\,km\,s$^{-1}$, {satisfying the photometric selections described in Sect.\,\ref{csel}}. \newline
Filled (green) circles indicate candidates selected in $g$, $r$ and $i$; open (blue) circles: candidates selected in one or two filters. 
Dots (magenta) represent R11 Hyades candidates (PS1 and SDSS photometry). 
The thick (magenta) line represents the BT-Settl model 600-Myr isochrone. } 
\label{CMDz}
\end{figure}

{We note that the  discrimination between the Hyades cluster sequence and the bulk of the background stars is the better the larger is the wavelength difference between the bands.} 
In particular, $z$ and $y$ photometry alone combined with 2MASS photometry does not provide a good rejection (see Fig.\,\ref{CMDz}). {We note again that the isochrone does not reproduce the observations.} We therefore concentrate on $g$, $r$ and $i$ bands in the following. 
For each CMD, we fit the Hyades stellar sequence with a second-order polynomial:

\begin{align*}
M_g = 9.628+0.8601\times(g-K_s)+0.04491\times(g-K_s)^2 \\
M_r =  9.834+1.1000\times(r-K_s)+0.05701\times(r-K_s)^2 \\
M_i = 10.312-1.4802\times(i-K_s)+0.07716\times(i-K_s)^2
\end{align*}

{These constraints are only valid for stars with $M_{K_s}>6$\,mag.
 Candidates presumably brighter than this are better selected in R11 and are not considered in this study.}
{To refine the selection box in order to minimize the contamination from field stars},  {we perform a numerical simulation of the field star contamination using the Besan\c{c}on model} (see Sect.\,\ref{model}). We adopt the following {selection criteria: we select candidates redder than the fitted Hyades sequence shifted by $-0.2$\,mag and bluer than the Hyades sequence of equal-luminosity binaries 
(see dotted lines in Fig.\,\ref{CMDs}).}
 {This would allow the selection of equally-bright binaries.}

Because of the partial sky coverage of those filters, we select candidates detected in any of those bands and satisfying our criterion. 
{We reject a candidate if it fails our criterium in any of the available CMDs.}
We select {273}~PS1 candidates objects with $|\rm{v}_\perp|<4\rm\,km\,s^{-1}$.
We perform the same analysis with SDSS--DR8 photometry, and select {30} SDSS candidates.
There are {29}\,candidates selected with both PS1 and SDSS photometry. The SDSS DR8 catalogue contributes {no} candidate, and rejects {no} PS1 candidate, while PS1 rejects {one} SDSS candidate. 
The sample holds {(again) 273} candidates.

{In what precedes we have used the median PSF magnitudes of the good PS1 measurements.  
We can compare our PSF magnitudes with the aperture magnitude, to remove problematic measurements, such as poorly resolved visual binaries, or extended objects. The proper motions of such objects would likely be unreliable. We decide to remove the 13 candidates with PSF and aperture magnitudes discrepant by more than 0.5\,mag. 
This cut only removes the few particularly poor measurements, and the dispersion of the magnitude differences of the remaining candidates is 0.05\,mag. 
Most {of the} removed candidates were selected in a single band, have $|\rm{v}_\perp|>2\rm\,km\,s^{-1}$, or are suspected to be giants (see below).
}

\subsection{Rejection of background giant stars and erroneous proper motions} \label{selgiants}

In Fig.\,\ref{JHHK}  we present the traditional near-infrared colour-colour diagram where giants tend to populate the upper-left side of the diagram above the sequence of the dwarf stars.
Again the isochrones do not describe accurately the observations.  
{Fig.\,\ref{JHHK} suggests a systematic offset of 0.04\,mag in the $H$ band.} 
We find {15}\,candidates that seem to display high extinction values. 
The excess maps of \citet{Schle98} indicate large amounts of dust on their sky location (see Fig.\,\ref{dust}), {mostly at low Galactic latitude or towards the Taurus-Auriga star forming region}.
{If truly reddened, these objects can only be located at large distances beyond the Hyades cluster. 
For instance, \citet{Tambu02}, in a study of 1000\,nearby stars, determines that the contribution to the polarization by the interstellar medium seems to become effective only after $\approx 70$\,pc, while our search excludes candidates beyond 77\,pc. (90\% of these objects have secular distances smaller than 70\,pc.)}  

{The space velocity of those objects whose near-infrared colours may indicate that they are giant stars, would be so large that they could} be members of the Hyades; or, more likely, the PPMXL proper motion is erroneous. 

We therefore reject candidates with:

\begin{align*}
J-H>&0.74 {\;\rm mag\; and}  \\
J-H>&0.74+1.6*(H-K_s-0.34) {\;\rm mag}
\end{align*}

In addition, we inspect the neighbourhood of the candidates in VizieR. While PPMXL used the individual epochs from the sources of USNO-B1.0 \citep{Roese10} to derive proper motions, it relies, however, on the cross-matching automatically done in USNO-B1.0. So, for the small number of candidates that survived kinematic and photometric selection, we perform a visual inspection on VizieR and on the digitised sky survey charts from IRSA (NASA/IPAC Infrared Science Archive).
{Through our visual inspection and the analysis presented in Sect.\,\ref{confastrom}, 
we remove {12} candidates that have a close neighbour or a likely erroneous proper motions, and keep {236}\,candidates.

\begin{figure}
\includegraphics[width=.5\textwidth]{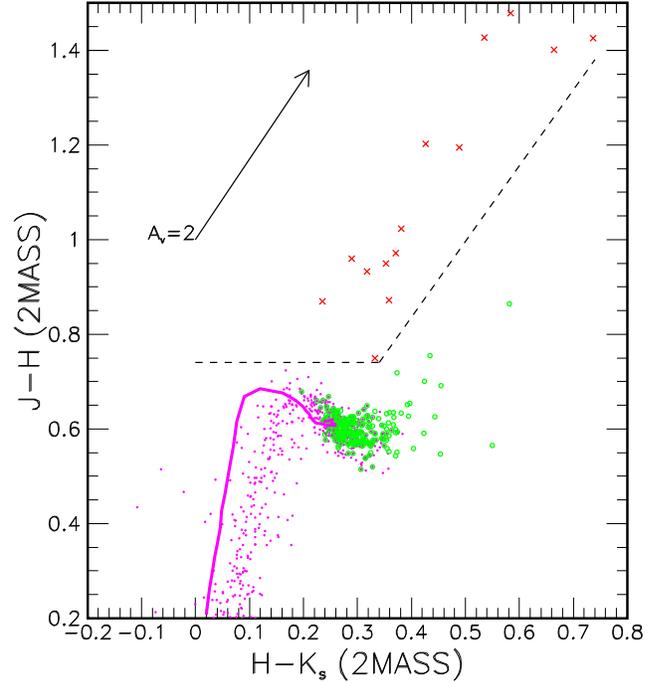}
\caption{Near-infrared colour-colour diagram of our candidates, after the selection in the optical-near-infrared CMDs. The R11 candidates appear as magenta dots, the new candidates as green circles, and red crosses show {kinematical candidates with suspiciously} high reddening. Objects located above the dashed lines are rejected as background stars (red crosses). The BT-Settl model 600-Myr isochrone down to 0.07\,M$_\odot$ is shown as the thick (magenta) line,  and the reddening vector ($R_V=3.1$) as the black arrow.
}
\label{JHHK}
\end{figure}

\begin{figure}
\includegraphics[width=.5\textwidth]{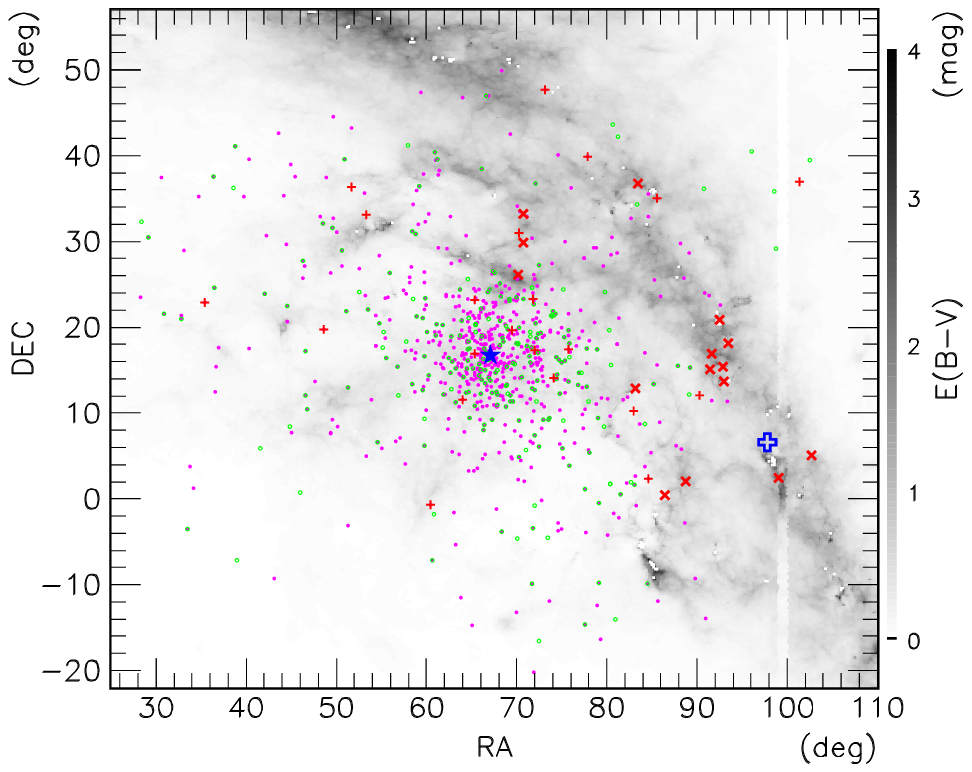}
\caption{Schlegel excess map of $E(B-V)$ \citep{Schle98}, with the darkest areas having the largest excess. The Hyades candidates are super-imposed: green circles for the candidates detected in PS1, with low reddening; red crosses and pluses: candidates suspected to be giants (see Fig.\,\ref{JHHK}) or to have bad proper motion measurements; magenta dots for R11.  The blue diamond shows the convergent point. }
\label{dust}
\end{figure}

The parameters available for each entry are described in Table\,\ref{CandParam}.
} 
 The full information, given in Table\,6 is available at the CDS. 
 Table\,\ref{CandParam} summarises its content: Columns\,1 and~2 list the J2000.0 positions as in PPMXL; 
 {Cols.\,3 and~4, the PPMXL identifier and flag which describes the quality of the proper motion fit and the origin of the measurement}; Cols.\,5 to~8, the PPMXL proper motions and uncertainties; Cols.\,9 and~10, the secular parallax and its uncertainty; Col.\,11, the distance to the cluster centre; Col.\,12, the predicted radial velocity; Cols.\,13 to~16, the tangential velocity in the direction and perpendicular to the direction to the convergent point; Col.\,17, the candidate mass; Cols.\,18 to~49, the PS1, SDSS, 2MASS, 
 and WISE magnitudes and their uncertainties; Col.\,50, the source of the candidate, discovered in R11 or in this paper.

\subsection{Further selection with UKIDSS and WISE photometry} \label{furthersel}

We use additional surveys to refine the selection. 
{UKIDSS observations are available for 92\,candidates, mostly in the $K$\,band. 
UKIDSS photometry confirms} all those candidates except one, \object{PPMXL~J63.4757+15.8357}, which has its $K$ magnitude discrepant with 2MASS's. 
It has two UKIDSS measurements, $K$=12.02 and $11.80$\,mag, 
both marked as close to saturated ({\tt k\_1ppErrBits={\tt 0x10}}), while 2MASS has $11.35$\,mag, 
with a negligible colour term.
There is no UKIDSS $ZYJH$ photometry. 
We ignore the UKIDSS measurements and keep the candidate.

\begin{figure*}
\includegraphics[width=.5\textwidth]{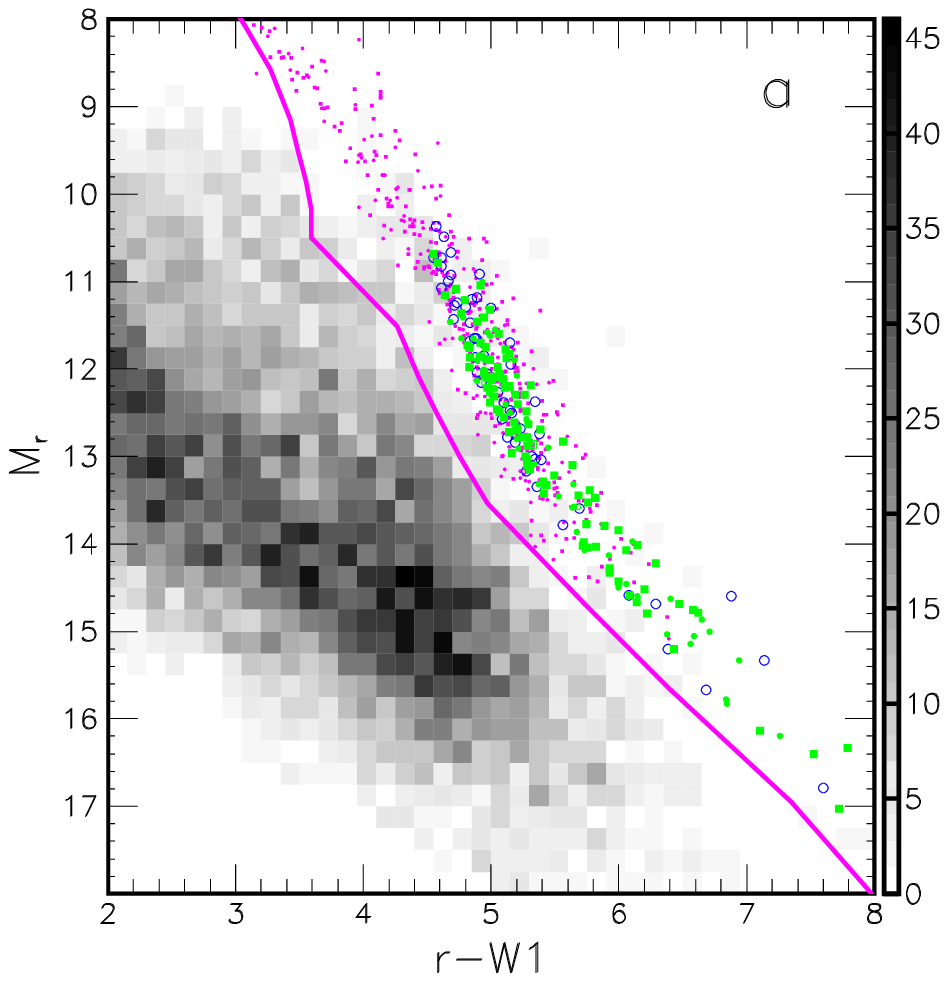}
\includegraphics[width=.5\textwidth]{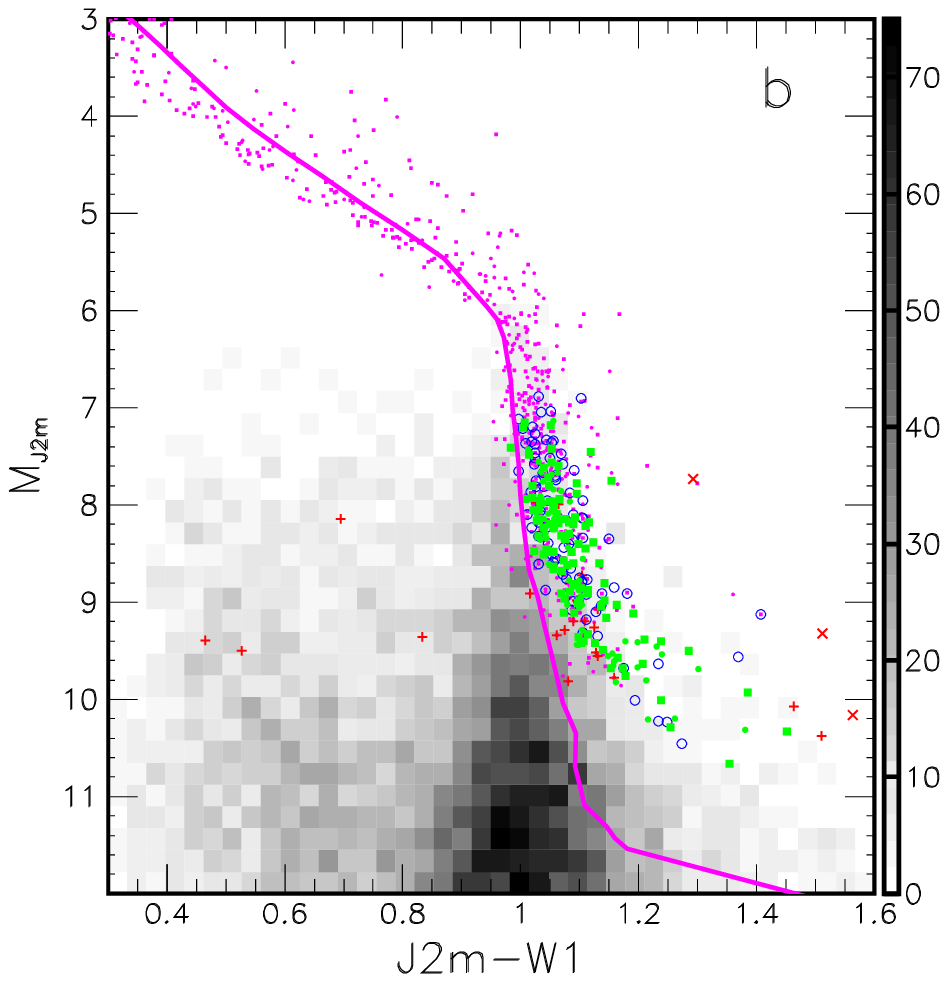}
\includegraphics[width=.5\textwidth]{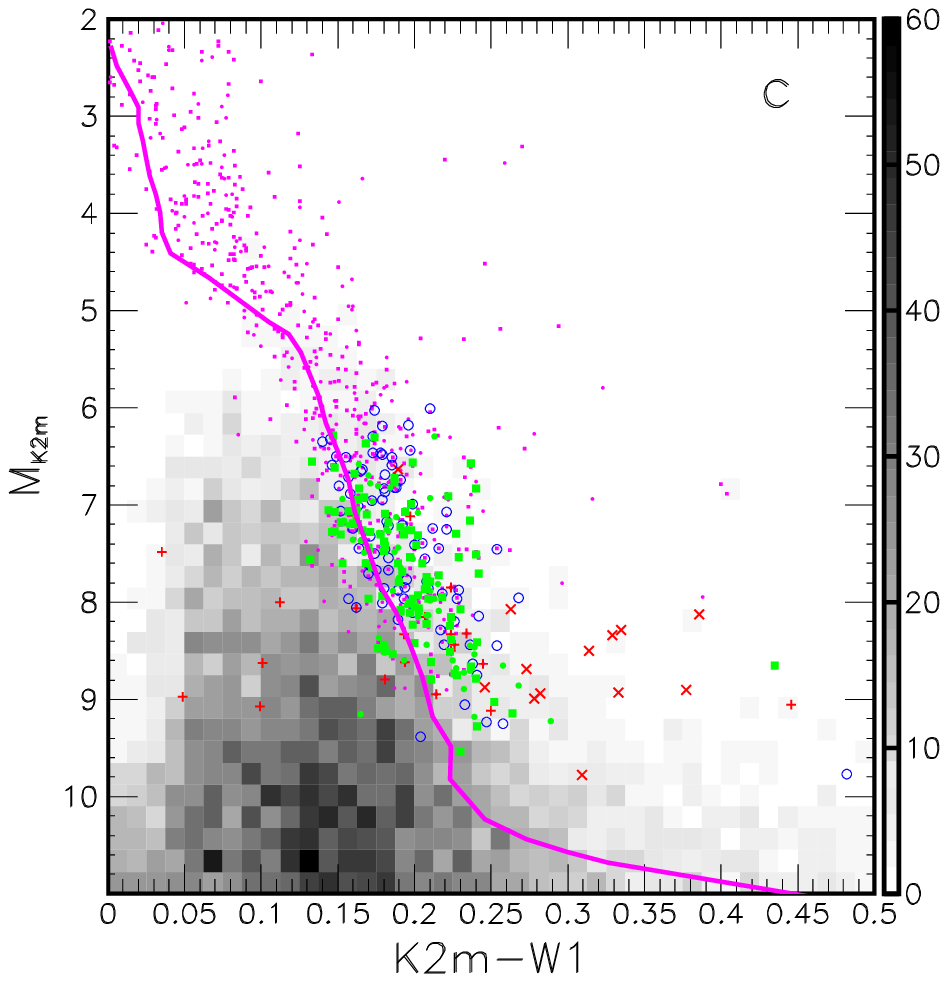}
\includegraphics[width=.5\textwidth]{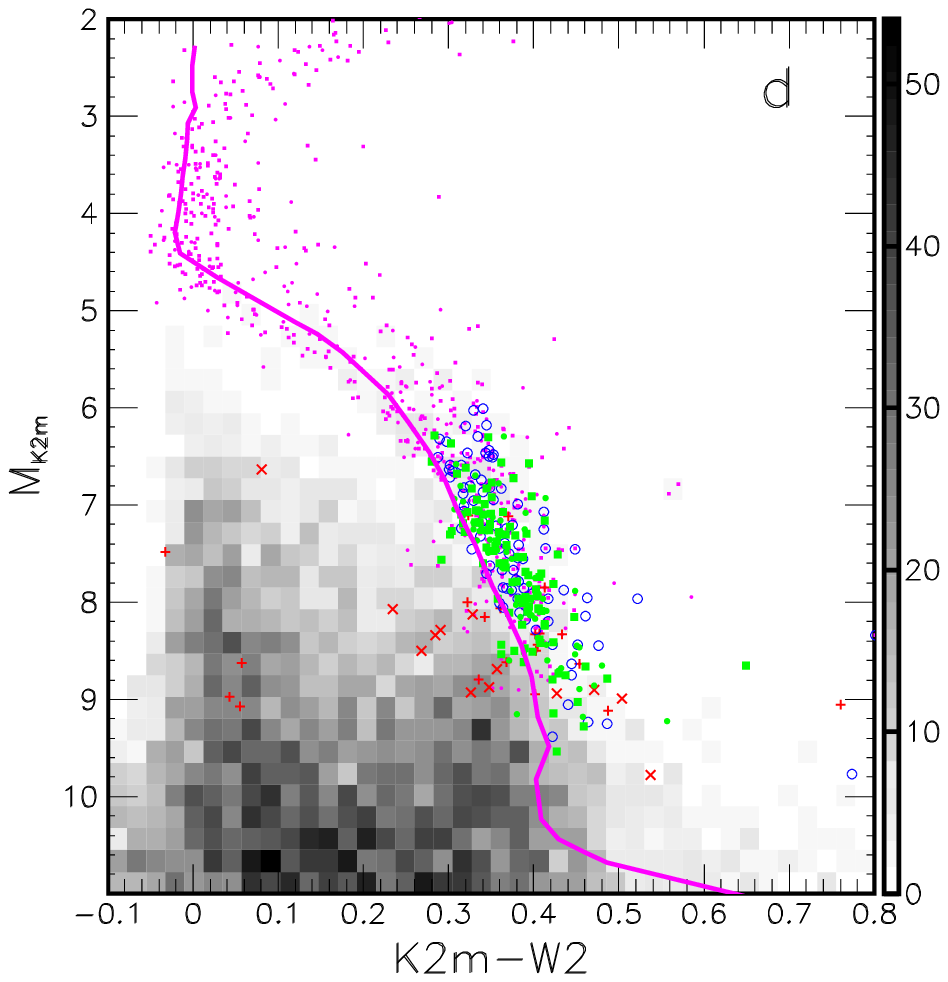}
\caption{
Colour-magnitude diagrams for the kinematic candidates with $|v_\bot|<4$\,km\,s$^{-1}$ within 30\,pc of the cluster centre: {\it a)} of $r_{\rm P1}-K_{\rm W1}$; {\it b)} of $J_{\rm 2MASS}-{\rm W1}$; {\it c)} of $K_{\rm 2MASS}-{\rm W1}$, and {\it d)} of $K_{\rm 2MASS}-{\rm W2}$. \newline
Filled (green) circles indicate candidates selected in $g$, $r$ and $i$; open (blue) circles: candidates selected in one or two filters; (red) crosses and pluses: candidates suspected to be giants (see Fig.\,\ref{JHHK}) or to have bad proper motion measurements. 
{\it b, c, d)}  
Dots (magenta) represent R11 Hyades candidates, with CMC14 $r$-band photometry in {\it c}. 
The thick (magenta) line represents the BT-Settl model 600-Myr isochrone. } 
\label{CMDIRs}
\end{figure*}

{We cross-matched our kinematic candidate catalogue with the WISE sky data release \citep{Wrigh10}, within 5\,arcsec of the J2011.0 PPMXL position. We recover all photometric candidates but a single $K=12.2$-mag candidate. 
We construct CMDs using optical and near-infrared (2MASS) photometry combined with WISE W1 and W2 bands (Fig.\,\ref{CMDIRs}).
Most candidates follow a sequence, again shifted from the theoretical isochrone, but only by +0.05\,mag in color.
{ Only one candidate appears much bluer than the bulk of the candidate,  \object{PPMXL~J101.3119+36.9271}. This candidate, which has a double entry in PPMXL, is detected in PS1 $r$ band only, and is located 28\,pc from the cluster centre.
 It is most probably a spurious candidate, and we remove it.} 
}

\begin{table}
 \centering
  \caption{Basic description of the parameters of the R11 candidates and the 62 new candidates (table available at the CDS).
  Errors are included for most parameters. See Vizier for a complete description and format of the file.}
  \begin{tabular}{rlll}
\hline
\hline
  Col. & Label & Unit & Description \\
\hline
1 & ID & & Running number (as in R11 if available) \\
2  &RA      &  deg   &  PPMXL Right Ascension\\
3  &DE      & deg   &  PPMXL  Declination\\
    &            &           & both J2000.0, epoch 2000.0\\
4  & RA\_PS1      &  deg   &  PS1 Right Ascension\\
5  & DE\_PS1      & deg   &  PS1  Declination\\
    &            &           & both J2000.0\\
6 & MJDps & d & epoch of PS1 positions  \\
7  & PPMXL\;ID &    & PPMXL identifier \\    
8  & PPMXL flag & & see \citet{Roese10} \\
9  &pmRA        & mas\,yr$^{-1}$ & Proper motion in RA$\cdot$cos(DE)\\
11  &pmDE        & mas\,yr$^{-1}$ & Proper motion in DE\\
13 &$\varpi$    & mas   &  Secular parallax\\
15 &$\rm{r_c}$    & pc  &  Distance from the cluster centre\\
16 &$\rm{v}_r$    & km\,s$^{-1}$  &  Predicted radial velocity\\
17 &$\rm{v}_\parallel$    & km\,s$^{-1}$  &  Tangential velocity in direction \\
     &    &       & to the convergent point\\
18 &$\rm{v}_\perp$    & km\,s$^{-1}$      &  Tangential velocity perpendicular to \\
   &    &       & the direction to the convergent point\\ 
21 &$m$    & $\rm{M}_\odot$        &  mass of the star \\
22 & $g_{\rm P1}$ & mag & PS1 $g$ {PSF median} magnitude \\
24 & $r_{\rm P1}$ & mag & PS1 $r$ {PSF median} magnitude \\
 26 & $i_{\rm P1}$ & mag & PS1 $i$ {PSF median} magnitude \\
 28 & $z_{\rm P1}$ & mag & PS1 $z$ {PSF median} magnitude \\
 30 & $y_{\rm P1}$ & mag & PS1 $y$ {PSF median} magnitude \\
 32 & $g'$ & mag & SDSS {DR8} $g$ magnitude \\
 34 & $r'$ & mag & SDSS {DR8} $r$ magnitude \\
 36 & $i'$ & mag & SDSS {DR8} $i$ magnitude \\
 38 & $z'$ & mag & SDSS {DR8} $z$ magnitude \\
 40 & $J_{\rm 2M}$ & mag & 2MASS $J$ magnitude \\
 42 & $H_{\rm 2M}$ & mag & 2MASS $H$ magnitude \\
 44 & $K_{s\rm 2M}$ & mag & 2MASS $K_s$ magnitude \\
 46 & $W1$ & mag & WISE $W1$ magnitude \\
 48 & $W2$ & mag & WISE $W2$ magnitude \\
 50 & $W3$ & mag & WISE $W3$ magnitude \\
 52 & $W4$ & mag & WISE $W4$ magnitude \\
 54 & Source flag &  & 0: R11; 1: this paper \\
\hline
\label{CandParam}
\end{tabular}
\end{table}

\subsection{Status of the R11 candidates}

We search the {724} candidates of R11 in our catalogues as we did for the kinematic sample as described above.
Many R11 candidates are bright and possibly saturated in PS1 or SDSS, so that we only consider candidates with {PS1 or SDSS} $g>16, r>16$ or $i>16$\,mag, which is conservative. The CMC14 photometry of brighter objects is more reliable than the PS1 and SDSS photometry.
 {We check the location of those 
 fainter R11 candidates in the colour-magnitude diagrams, as described in Sect.\,\ref{optselection}, using SDSS and/or PS1 photometry. }
{Seven} candidates are rejected in one or more colour-magnitude diagrams (see Table\,\ref{R11phrejec}) and we remove them from the subsequent analysis.

{We also check the proper motions of the fainter R11 candidates following the analysis presented in Sect.\,\ref{confastrom}. We find that six~candidates have offsets in the PS1 catalogue of 300\,mas or more (see Table\,\ref{R11astrejec}), and we remove them from the sample.
}

\subsection{Final sample}

Our PS1+SDSS sample is made up of {236}~candidates, with $|\rm{v}_\perp|<4\rm\,km\,s^{-1}$.
 Among those, {62} candidates were not included in the R11 candidates. 
We provide the PS1 photometry and a subset of relevant parameters of these new candidates in Table\,\ref{NewCand}. 

 {Among those {62} candidates, eight were previously identified as Hyades candidates (see Sect.\,\ref{inters}).}
{Turning now to the ``best'' candidates with $|\rm{v}_\perp|<2\rm\,km\,s^{-1}$, we select {207}~candidates, with {48} not included in the 620 remaining R11 candidates having $|\rm{v}_\perp|<2\rm\,km\,s^{-1}$. 
We combine our two samples into a list of  668 cluster member candidates.}
{The proper motions of R11 differs slightly from the PPMXL proper motions because the astrometry of CMC14 was combined with the PPMXL measurements. Whenever available we use the proper motions of R11.}

\begin{table*}
 \centering
 \begin{minipage}{\textwidth}
  \caption{Optical and 2MASS photometry of the R11 candidates rejected in our study because of their photometry.
    }
  \begin{tabular}{r@{}lr@{$\pm$}lr@{$\pm$}lr@{$\pm$}lr@{$\pm$}lr@{$\pm$}lcr@{$\pm$}lr@{$\pm$}lr@{$\pm$}lcc}
  \hline
  \hline
   \multicolumn{2}{l}{Name}     & \multicolumn{2}{c}{ $g_{\rm P1}$ or $g$}& \multicolumn{2}{c}{$r_{\rm P1}$ or $r$} & \multicolumn{2}{c}{$i_{\rm P1}$ or $i$} & \multicolumn{2}{c}{$z_{\rm P1}$ or $z$}  &  \multicolumn{2}{c}{$y_{\rm P1}$} & surv. & \multicolumn{2}{c}{$J$} & \multicolumn{2}{c}{$H$} & \multicolumn{2}{c}{$K$}  \\
   \multicolumn{2}{l}{(deg, J2000.0)}     & \multicolumn{2}{c}{(mag)}& \multicolumn{2}{c}{(mag)} & \multicolumn{2}{c}{(mag)} & \multicolumn{2}{c}{(mag)}  &  \multicolumn{2}{c}{(mag)} & & \multicolumn{2}{c}{(mag)} & \multicolumn{2}{c}{(mag)} & \multicolumn{2}{c}{(mag)}  \\
 \hline
35.415 & +22.851 & 18.06 & 0.01& 16.77 & 0.01& \multicolumn{2}{c}{---} & 14.27 & 0.00& 13.89 & 0.01& PS1 & 12.57 & 0.02 & 11.96 & 0.02 & 11.64 & 0.02 \\
 & &  18.14 & 0.01 &  16.87 & 0.01 &  15.10 & 0.00 &  14.30 & 0.00 &  \multicolumn{2}{c}{---} & SDSS \\
 & &  \multicolumn{2}{c}{---} &  16.76 & 0.24 &  \multicolumn{2}{c}{---} &  \multicolumn{2}{c}{---} &  \multicolumn{2}{c}{---} & CMC \\
60.394 & -0.726 & \multicolumn{2}{c}{---} & 16.84 & 0.01& \multicolumn{2}{c}{---} & 14.10 & 0.01& 13.69 & 0.01& PS1 & 12.35 & 0.02 & 11.75 & 0.02 & 11.48 & 0.03 \\
 & &  \multicolumn{2}{c}{---} &  16.92 & 0.24 &  \multicolumn{2}{c}{---} &  \multicolumn{2}{c}{---} &  \multicolumn{2}{c}{---} & CMC \\
65.320 & +23.198 & 17.37 & 0.01& 16.08 & 0.01& 14.61 & 0.01& 13.71 & 0.01& 13.33 & 0.01& PS1 & 12.12 & 0.02 & 11.51 & 0.02 & 11.24 & 0.02 \\
 & &  17.27 & 0.01 &  16.06 & 0.00 &  14.47 & 0.00 &  13.70 & 0.00 &  \multicolumn{2}{c}{---} & SDSS \\
 & &  \multicolumn{2}{c}{---} &  16.19 & 0.24 &  \multicolumn{2}{c}{---} &  \multicolumn{2}{c}{---} &  \multicolumn{2}{c}{---} & CMC \\
48.638 & +19.720 & 17.49 & 0.01& 16.17 & 0.01& 14.58 & 0.01& 13.82 & 0.01& 13.50 & 0.01& PS1 & 12.19 & 0.02 & 11.64 & 0.03 & 11.34 & 0.02 \\
 & &  17.51 & 0.01 &  16.25 & 0.00 &  14.61 & 0.00 &  13.86 & 0.00 &  \multicolumn{2}{c}{---} & SDSS \\
 & &  \multicolumn{2}{c}{---} &  16.45 & 0.24 &  \multicolumn{2}{c}{---} &  \multicolumn{2}{c}{---} &  \multicolumn{2}{c}{---} & CMC \\
73.164 & +47.698 & \multicolumn{2}{c}{---} & 17.05 & 0.01& \multicolumn{2}{c}{---} & 14.54 & 0.01& 14.19 & 0.00& PS1 & 12.83 & 0.02 & 12.20 & 0.02 & 11.86 & 0.03 \\
 & &  \multicolumn{2}{c}{---} &  17.44 & 0.24 &  \multicolumn{2}{c}{---} &  \multicolumn{2}{c}{---} &  \multicolumn{2}{c}{---} & CMC \\
84.599 & +2.402 & 18.10 & 0.01& 16.82 & 0.01& \multicolumn{2}{c}{---} & 14.33 & 0.01& 13.96 & 0.01& PS1 & 12.61 & 0.02 & 12.05 & 0.03 & 11.76 & 0.03 \\
 & &  \multicolumn{2}{c}{---} &  16.84 & 0.24 &  \multicolumn{2}{c}{---} &  \multicolumn{2}{c}{---} &  \multicolumn{2}{c}{---} & CMC \\
85.522 & +34.978 & 17.80 & 0.01& 16.52 & 0.01& \multicolumn{2}{c}{---} & 14.39 & 0.01& 13.78 & 0.01& PS1 & 12.51 & 0.02 & 11.93 & 0.02 & 11.63 & 0.02 \\
 & &  \multicolumn{2}{c}{---} &  16.80 & 0.24 &  \multicolumn{2}{c}{---} &  \multicolumn{2}{c}{---} &  \multicolumn{2}{c}{---} & CMC \\
 \hline
\label{R11phrejec}
\end{tabular}
\end{minipage}
\end{table*}

\begin{table*}
 \centering
 \begin{minipage}{\textwidth}
  \caption{PS1 photometry, PPMXL astrometry and PS1 offsets of the R11 candidates rejected in our study because of their astrometry. 
  }
  \begin{tabular}{r@{}lr@{$\pm$}lr@{$\pm$}lr@{$\pm$}lr@{$\pm$}lr@{$\pm$}lr@{$\pm$}lr@{$\pm$}lcc}
  \hline
  \hline
   \multicolumn{2}{l}{Name}     & \multicolumn{2}{c}{ $g_{\rm P1}$ or $g$}& \multicolumn{2}{c}{$r_{\rm P1}$ or $r$} & \multicolumn{2}{c}{$i_{\rm P1}$ or $i$} & \multicolumn{2}{c}{$z_{\rm P1}$ or $z$}  &  \multicolumn{2}{c}{$y_{\rm P1}$}  & \multicolumn{2}{c}{ $\mu_\alpha.\cos(\delta)$} &  \multicolumn{2}{c}{$\mu_\delta$} & $\Delta \alpha$ & $\Delta \delta$ \\
   \multicolumn{2}{l}{(deg, J2000.0)}     & \multicolumn{2}{c}{(mag)}& \multicolumn{2}{c}{(mag)} & \multicolumn{2}{c}{(mag)} & \multicolumn{2}{c}{(mag)}  &  \multicolumn{2}{c}{(mag)} & \multicolumn{2}{c}{(mas\,yr$^{-1}$)} & \multicolumn{2}{c}{(mas\,yr$^{-1}$)} & (mas) & (mas)\\
 \hline
64.045 & +11.534 & 18.23 & 0.01& 16.87 & 0.01& 15.08 & 0.01& 14.24 & 0.01& 13.95 & 0.01& +119.3 &  3.7 &   -2.2 &  3.6 & +211 & -292 \\
74.096 & +14.101 & 18.72 & 0.01& \multicolumn{2}{c}{---} & 15.71 & 0.00& 14.86 & 0.01& 14.44 & 0.00&  +75.3 &  4.0 &  -11.3 &  3.8 & -240 & -401 \\
65.398 & +16.894 & 16.11 & 0.01& 14.81 & 0.00& \multicolumn{2}{c}{---} & 12.45 & 0.01& 12.22 & 0.00& +130.2 &  4.5 &  -37.3 &  4.5 & -584 & +267 \\
75.715 & +17.451 & 17.92 & 0.01& 16.61 & 0.00& 14.59 & 0.00& 13.62 & 0.00& 13.14 & 0.01& +115.6 &  5.3 &  -38.3 &  5.1 & -487 &  +98 \\
71.754 & +23.263 & 18.13 & 0.01& 16.82 & 0.01& 15.05 & 0.01& 15.01 & 0.01& 13.86 & 0.01&  +95.8 &  3.9 &  -58.1 &  3.7 &  -97 & +290 \\
77.825 & +39.908 & 18.39 & 0.01& 17.14 & 0.01& 15.47 & 0.01& 14.77 & 0.01& 14.33 & 0.00&  +57.6 &  4.8 &  -75.1 &  4.7 & -235 & +469 \\
 \hline
\label{R11astrejec}
\end{tabular}
\end{minipage}
\end{table*}

\section{Objects of special interest} \label{inters}

 {We search our sample for known objects in Simbad and Vizier (all catalogues), within 5\arcsec\ of the epoch-2000.0 PPMXL position. We cannot discuss all observations of our candidates, so we note only those referring to cluster membership, spectral classification, age, or distance. }
The present study, based on the optical survey of USNO-B and the optical bands of PS1 is not very sensitive to brown dwarfs, except the closest ones.
Indeed we find none of the candidates of  \citet{Hogan08} in our final list.

 {Eleven candidates are listed as field dwarfs in \citet{Slesn06} (their Table\,A.1), identified from their Na--8189 index indicating an age $\gtrsim 100$\,Myr. Our candidates have an index between 0.78 and 0.82 typical of an intermediate age. }
  
\subsection{Hyades members from \citet{Bouvi08}}

We recover in our final list 13 of the candidates of  \citet{Bouvi08}: CFHT-Hy-1--6, 8--10, 12, 13, 15, and 17. 
 {Eleven of those were already found in R11; here we add CFHT-Hy-12 and 13. 
We do not select the remaining candidates of \citet{Bouvi08} for the following reasons: }
CFHT-Hy-11 is kinematically selected {but photometrically rejected in both the $M_K$ vs. $g_{\rm P1}-K_s$ and  $M_K$ vs. $i_{\rm P1}-K_s$ diagrams as too red by 0.04 and 0.09\,mag respectively. With $\rm{v}_\perp=2.2\rm\,km\,s^{-1}$, it seems likely to be a contaminant.}
Three candidates are not in PPMXL (CFHT-Hy-16, 20 and 21), while the five remaining candidates were rejected on the basis of their kinematics, because of discrepant proper motions between \citet{Bouvi08} and PPMXL.
{The proper motion of PPMXL for CFHT-Hy-12 and CFHT-Hy-13 differs from the proper motion of \citet{Bouvi08} by between 10 and $24\rm\,mas\,yr^{-1}$, but the PPMXL baseline is much longer. }
We  {note that we do not} select {objects from their Table\,D.1 (Hyades probable non-members)}.

 {
\subsection{Other known Hyades candidates}

We identify {six} of the previously known Hyades candidates in our final list {(ordered by RA)}: }

\begin{itemize}
\item \object{DENIS J0235495-071121} was classified as an M5.5 by \citet{PhanB06}, with a spectrophotometric distance of 25.5\,pc, while we predict a secular distance of 28.3\,pc. \citet{West08} classify it as an M7 star based on SDSS spectroscopy. \citet{Galvez10} assigns it a distance of $28.2\pm3.4$\,pc and a Hyades membership. {We predict a somewhat large perpendicular velocity of $(3.0\pm 0.8)\rm\,km\,s^{-1}$.}
{\item \object{Cl* Melotte 25 VA 200} (\object{LP 475-7}) is listed in \citet{Gicla62} as a Hyades member. 
 We derive $\rm{v}_\perp=(-3.8\pm 0.8)\rm\,km\,s^{-1}$.}
\item \object{LP 415-19} (\object{PPMXL J65.43489+20.40293}) was identified as a possible Hyades member by \citet{Reid92,Reid93}. It is unresolved by \citet{Reid00h}, but could be a double-lined system. The measured radial velocity, $36.82\pm2.10$\,km\,s$^{-1}$, again agrees with our expected value of 37.5\,km\,s$^{-1}$ 
{\item \object{PPMXL J65.45657+19.48572} (\object{LP 415-20}) has been classified as M6.5e Hyades member by \citet{Bryja94} . It is reported as an M7+M9.5 binary with a separation of 0.1\arcsec\ by \citet{Siegl03}.  \citet{Dupuy11} predicts a period of $14.4\pm0.4$\,years, so it will be possible to measure its mass. It is detected by XMM \citep{Watso09}.}
\item \object{PPMXL J66.57934+17.05058} (\object{LP 415-881}) was identified as a possible Hyades member by \citet{Reid92}.  It is unresolved by \citet{Reid00h}, {using Keck\,I/HIRES \'echelle high-resolution spectroscopy to identify spectroscopic binaries.} {We derive $\rm{v}_\perp=(-2.3\pm 1.1)\rm\,km\,s^{-1}$.} 
\item \object{PPMXL J69.74372+15.66145} (\object{LP 415-2128)} was identified as a possible Hyades member by \citet{Bryja92,Reid92}. It is unresolved by \citet{Reid00h}. The measured radial velocity, $39.5\pm0.5$\,km\,s$^{-1}$, agrees well with our expected value of 39.7\,km\,s$^{-1}$. 
\end{itemize}

A number of our member candidates have been identified as cool nearby stars, and have published spectral types obtained with spectroscopic observations. All those objects are classified as mid-M- to early-L-type low-mass stars, sometimes with hint of youth or binarity. {These are described in the next two sections.} 

\subsection{L-type dwarfs from Cruz et al. (2003,2007)}

We identify in our candidate sample two objects published as field L-type dwarfs:
\begin{itemize}
\item \object{2MASSI J0230155+270406} is classified as a peculiar L0: dwarf, with a spectrophotometric distance of $32.0\pm3.9$\,pc by \citet{Cruz07}.  {From the convergent-point analysis, } we estimate its parallax to be $22.5\pm0.6$\,mas, or $44.4\pm1.2$\,pc. {This is only 3-$\sigma$ away from the spectrophotometric distance, and we keep it as a candidate. }
\item \object{2MASS J05233822$-$1403022} is classified as a L2.5 dwarf in the optical (L5 in the near-infrared), with a spectrophotometric distance of $13.4\pm 1.1$\,pc by \citet{Wilso03} and \citet{Cruz03}. We estimate its parallax to be \mbox{$42.3\pm1.3$\,mas}, or \mbox{$23.6\pm0.8$\,pc}. 
Its spectrum shows signs of lithium \citep{Reine08}, and variable signs of X-ray, radio and H$\alpha$ emissions \citep[][and ref. therein]{Berge10}. No companion was found with separation larger then 0.15\arcsec\ by \citet{Reid08}.

\citet{Seifa10} and \citet{Blake10} measure a radial velocity of $+11.3\pm 1.0\,$km\,s$^{-1}$ and $+12.21\pm 0.09$\,km\,s$^{-1}$, while our kinematic analysis predicts a velocity of  $+40.6\,$km\,s$^{-1}$. 
Therefore its Hyades membership is questionable.
\end{itemize}

{
\subsection{M-type dwarfs with spectroscopic classification}

\begin{itemize}
\item \object{LP 245-52}  was first identified as a cool star by \citet{Cruz02} and finally classified as a M6 dwarf by  \citet{Cruz03}.  
\item \object{2MASS J03040207+0045512} (\object{DENIS	J030402.0+004551}) was first identified as a late-type star by \citet{Tinne93}, and classified as an M5.5 star by \citet{Bocha05}. Their corresponding spectrophotometric distance is 31\,pc, which agrees with our predicted distance of 30.0\,pc.
\item \object{2MASS J03300506+2405281} (\object{LP 356-770}) was classified as an M7 star by \citet{Gizis00}, with a spectrophotometric distance of 20\,pc, smaller than our expected value of 37\,pc. The radial velocity of $39.2\pm1.1$\,km\,s$^{-1}$ \citep{Reid02h} agrees marginally with our expected value of 31.1\,km\,s$^{-1}$.
\item \object{2MASS J03540135+2316339} has been classified as M8 dwarf by \citet{Kirkp97}. \citet{Cruz03} estimates a spectrophotometric distance of $22.9\pm1.7$\,pc for a spectral type of M8.5, again smaller than our expected value of 31.05\,pc. It is not resolved into a binary (with separation larger then 0.10\arcsec) by \citet{Close03}. It is also detected by XMM \citep{Watso09}.
\item \object{PPMXL J57.73917+18.30189} (\object{LP 413-53}, \object{2MASS J03505737+1818069}) was identified as an M9 star by \citet{Gizis00}. Various spectrophotometric distances have been proposed, close to $19.9$\,pc \citep{Gizis00}, smaller (by about $3\sigma$) than our prediction of 33.7\,pc.  {\citet{Reid02h} observed the object when it was undergoing a strong flare. They measure and discuss its radial velocity: it is significant smaller (either $5.8\pm2.0$\,km\,s$^{-1}$ or $-14.3\pm0.4$\,km\,s$^{-1}$ depending on the method) than our prediction of 34.4\,km\,s$^{-1}$. However the unusual H$\alpha$ profile, which may cause the discrepant measurements, is reminiscent of their observation of the Hyades SB2 binary. }
\item \object{2MASSI J0411063+124748} was identified as an M6 star by \citet{Cruz07}.
{\item \object{2MASS J04273708+2056389} is listed as a M5.75 field dwarf by \citet{Luhma06sd}, based on youth indicators of the optical and IR spectra, meaning that it is older than a few Myr. 
\item \object{2MASS J04441479+0543573} is listed as a M8 dwarf by \citet{Reid08X}.
\citet{Faher09} gives a proper motion of $\mu_(\alpha,\delta)=(+95,-6)\pm 21\rm\,mas\,yr^{-1}$ in agreement with the more accurate PPMXL measurement of $\mu_(\alpha,\delta)=(+77.01,+0.73)\pm 5.3\rm\,mas\,yr^{-1}$. 
}
\end{itemize}

\subsection{Spectrophotometric vs. secular parallaxes}

For the candidates with spectroscopic classification, we can compare the spectrophotometric distance with our predicted secular distance. In Table\,\ref{TabDist}, we list {our secular distance, the distance we calculate based on the 2MASS $J$-band magnitude, using the $M_J$ magnitudes given by \citet{West05}, and the published distance from the literature}. Deriving the distance from the spectral type and observed magnitudes, assuming a relationship based on old, field stars, is not an accurate process, given the spread of absolute magnitudes for a given spectral type. For most of the stars in Table\,\ref{TabDist}, the distances differ, marginally for any object, but the spectrophotometric distance seems systematically underestimated.

Previous studies did not notice such large differences {between the Hyades cluster members and the older field population}:
The latest BT-Settl models predict a radius of $R=0.116\,\rm R_\odot$ at 600\,Myr for a mass of $m=0.09\,\rm M_\odot$, and $R=0.112\,\rm R_\odot$ for ages of 2\,Gyr and older \citep{Allar11}. 
\citet{Reid00h} compared their secular parallaxes with the spectrophotometric distances cited in \citet{Reid93}, typically finding them 5\% larger. 
\citet{Siegl03} did not notice any difference between the Hyades and field CMDs, within their uncertainties (their Sect.\,4.2).

The differences in the distances may be caused by unresolved binaries.
{Alternatively, they could be field stars, although in this case}, we would also expect some candidates with an underestimated secular distance.
{We see none, but the sample size is small.}

{The low-mass stars of the Hyades may still be in the contraction phase, while the field population has settled on the main sequence.} 
However this low-significance effect needs to be confirmed by our on-going parallax measurements with PS1 \citep[see also][]{Andre11}.
}

\begin{table*}
 \centering
 \begin{minipage}{\textwidth}
  \caption{Published spectrophotometric and secular distances for a subset of the candidates with spectroscopic classification. 
  }
  \begin{tabular}{llcccl}
  \hline
  \hline
   Name                  & Sp.T.          & References     & Published   &  \citet{West05} & Secular \\
\hline
2M J0304+0045   & M5.5         & \citet{Bocha05} & 31                     & 25.1 & 30.0     \\ 
DE J0304+0045   & M5.5         & \citet{PhanB06} & 25.5                  & 33.7 & 28.3   \\ 
                                 & M7            & \citet{West08}   & 25.5                  & 20.2 & 28.3   \\ 
LP 415-20              & M7+M9.5 & \citet{Siegl03}   & $30\pm5$       & 27.2 & 40.2  \\ 
LP 413-53              & M9            & \citet{Gizis00}   & 19.9                  & 20.3 & 33.7  \\ 
2M J0230+2704   & L0:            & \citet{Cruz07}   & $32.0\pm3.9$  & 28.7 & 44.4  \\ 
2M J0523$-$1403 & L2.5        & \citet{Cruz03}   & $13.4\pm 1.1$ & 13.3 & 23.6  \\ 
\hline
\label{TabDist}
\end{tabular}
\end{minipage}
\end{table*}

\section{Luminosity function} \label{lf}

\subsection{Detection efficiency} \label{PPMXLdepth}

We performed an analysis similar to that of PS1 in Sect.\,\ref{PSdepth} for the PPMXL catalogue.

\begin{figure}
\includegraphics[width=.5\textwidth]{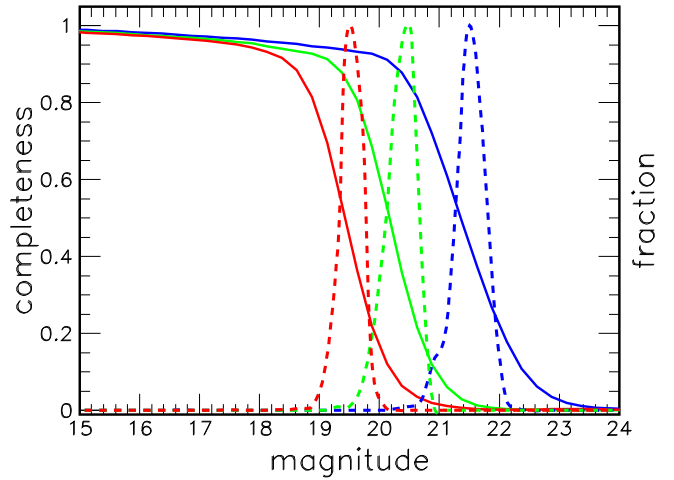}
\caption{Completeness of PPMXL as a function of $g$ (blue), $r$ (green), and $i$ (red) SDSS magnitudes (solid lines, right to left), and histograms of the 50\% completeness limit (dashed curves, normalized to unity).}
\label{depthPPMXL}
\end{figure}

We find that the PPMXL has a 50\% completeness limit of $g=21.5$, $r=20.3$ and $i=19.5$\,mag, with respective widths of 0.24, 0.22 and 0.18\,mag (see Fig.\,\ref{depthPPMXL}).
Because PPMXL requires multiple detections in various filters, and also because the USNO-B1.0 photometric system differs from that of PS1 and SDSS, the detection efficiency is a function of the object colour. 
 {We find significant incompleteness for the faintest and the reddest objects}.  
Therefore they are typically undetected on the blue plates, and detections on at least one red plate is required. Hence we calculate the completeness based on the $r$-band magnitude, as the $r$ filter profile is the closest to the red plate emulsion sensitivity profile. 
When that magnitude is unavailable, we extrapolate it from the $i$ or $g$ magnitudes, using the colours of our other Hyades candidates. 
In practice, the correction is smaller than a factor of 2 for all our candidates with a mass larger than $0.1\,\rm M_\odot$. 

{Because of our image examination by eye, because of the limit on the positional offsets (see Sect.\,\ref{confastrom}), or because of the rejection of candidates with giant-like near-infrared colours and/or dust reddening, it is possible that a few genuine members are rejected. }
We do not try to estimate the possible incompleteness due to these selections.
 {This would require a better understanding of the systematics in PPMXL, and an accurate calculation would require a detailed error model which is beyond the scope of this paper.}

\subsection{Contamination by field stars} \label{model}

We run the Besan\c{c}on model simulation to create a catalogue of {``mock''} stars with kinematics and Johnson-Cousins $BVRIK$ photometry \citep{Robin03}.
We run the model three times over the whole {field covered by our search (Fig.\ref{PS1cov}). 
Despite this, the Poisson noise is large at the end of the main sequence at which a small number of fake candidates is selected. }
We randomize the positions, as the model is calculated over a coarse grid.
We transform the Johnson-Cousins photometry into the $gri$ SDSS system according to \citet{Jordi06}. Finally we use the PS1 recovery fraction determined in Sect.\,\ref{Sec:skycov} and the PPMXL completeness (in SDSS $r$ band, see Sect.\,\ref{PPMXLdepth}) to randomly decide whether a ``fake'' star has $g$, $r$ and $i$ detections or not. 

\begin{figure}
\includegraphics[width=.5\textwidth]{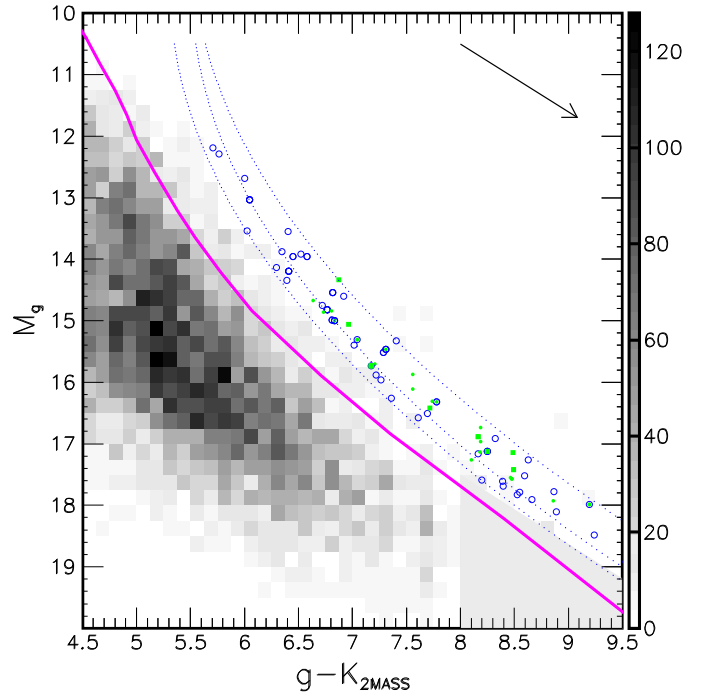}
\caption{Colour-magnitude diagram of our  ``fake'' kinematic candidates. 
As in Fig.\,\ref{CMDs} {\it b}, filled circles indicate candidates selected in $g$, $r$ and $i$; open circles: candidates selected in one or two filters; triangles: objects selected in $g$ but rejected in $r$ or~$i$.
The statistics is threefold the actual statistics, as we plot three realizations of the Besan\c{c}on model. 
}
\label{CMDsim}
\end{figure}

We then run our kinematic and photometric selection to obtain a synthetic catalogue of field contaminants (Fig.\,\ref{CMDsim}).
We find that almost all (92\%) contaminants are dwarfs, and {the remaining}, giants.

\begin{figure}
\includegraphics[width=.5\textwidth]{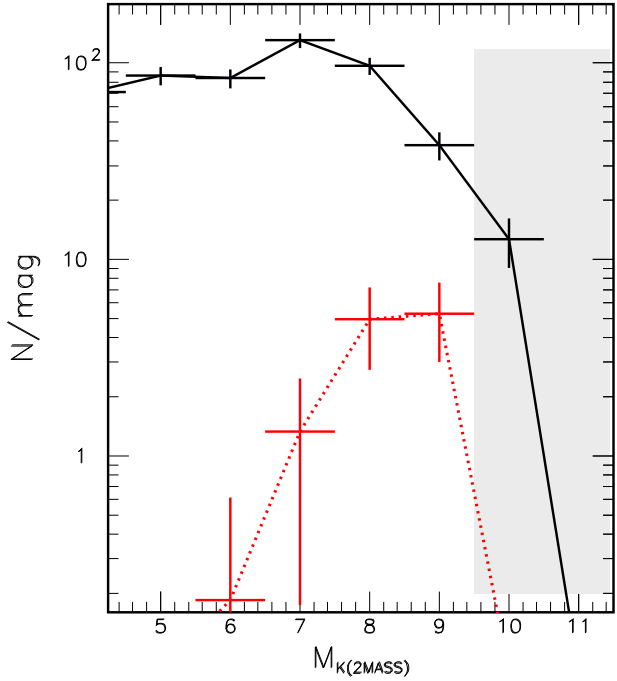}
\caption{
Luminosity function of the observed Hyades candidates (solid line) and of the simulated field contaminants (dotted line), for candidates with $|\rm{v}_\perp|<2\rm\,km\,s^{-1}$ {and outside the cluster core, i.e. with a distance to the cluster centre between 3.1 and 30\,pc}.
 The light grey area for $M_K>9.5$\,mag indicates the magnitudes for which our results are strongly affected by incompleteness. 
}
\label{LFcorr}
\end{figure}

We find that the contamination by field stars is negligible {($<10$\%) in cluster shells up to 18\,pc} of the cluster centre (see Fig.\,\ref{LFcorr}).
R11 considered actual observations of bright stars and the model of \citet{Kharc96} and reached the same conclusion. 
At larger distances from the centre, the cluster member density decreases and the survey volume increases, and with it the field contamination. 
In these outer areas, {for distances to the cluster centre between 18 and 30\,pc,} we find that the contamination increases up,  {for instance to about 17\% for candidates with $7.5<M_K<9.5$\,mag}.
{We stress that the precision of the simulation is rather poor, as the prediction of the Besan\c{c}on model over such a large area and range of Galactic latitudes is inaccurate. }
The effect on the luminosity function, however, is  small, especially at the very faint end, when Poisson noise and the effect of completeness corrections dominate the error budget.

\subsection{Corrected luminosity function}

We derive the cluster luminosity function over various cluster radii, to study its spatial dependence.
The {core} of the cluster is considered to a radius of 3.1\,pc.
We then consider the volume up to the tidal radius $r_t$=9\,pc (R11).
The halo is defined between the tidal radius and $2 r_t=18$\,pc.
Finally, the outer volume of our study covers a region between $2 r_t$ and 30\,pc.

We first correct the observed luminosity function for the detection efficiency: 
for each candidate, we obtain the PPMXL detection efficiency corresponding to its PS1 or SDSS {$r$-band magnitude, as described above. }
Here we consider the detection efficiency, as determined in Sect.\,\ref{PPMXLdepth}. 
For the {62}~Êcandidates found in PS1 or SDSS only, we also include the PS1+SDSS coverage probability.
We then give the candidates a weight inverse of the combined detection efficiency.

For each cluster radius range, we then subtract the contamination derived from the simulation,  corrected for incompleteness as the candidate sample. 
The correction of the PPMXL incompleteness is the largest effect at faint magnitudes, and our results for $M_K>9.5$\,mag are very sensitive to the details of the procedure, and therefore not solid. 
The second largest effect is the correction of the field contamination, which mostly increases with the distance to the cluster centre, here up to 30\% (for $M_K=9$\,mag). 

\begin{figure}
\includegraphics[width=.5\textwidth]{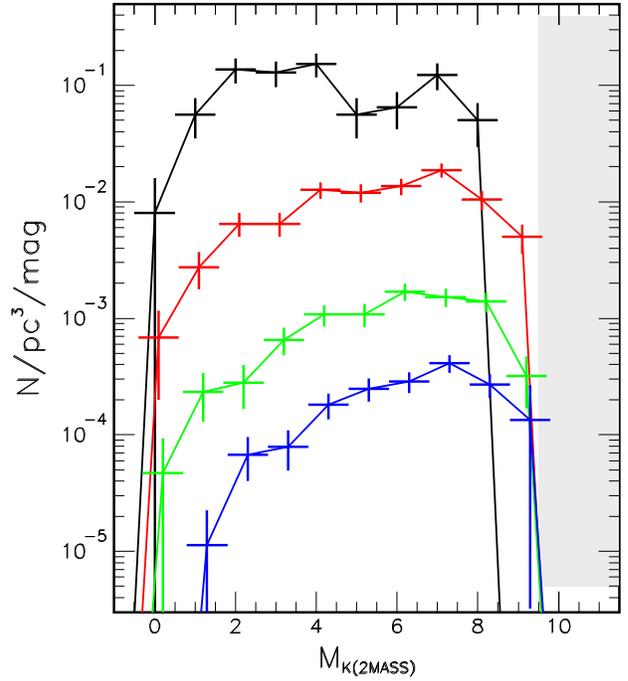}
\caption{Corrected luminosity function (density of candidates {with $|\rm{v}_\perp|<2\rm\,km\,s^{-1}$}) of the Hyades in the 2MASS $K_s$ band, for regions of various cluster radii: 0 to 3.1 \,pc; 3.1 to 9\,pc; 9 to 18 \,pc; and 18 to 30 \,pc (top to bottom). 
The light grey area for $M_K>9.5$\,mag indicates the magnitudes for which our results are strongly affected by incompleteness. 
}
\label{LF}
\end{figure}

The result is shown in Fig.\,\ref{LF} for four different shells around the cluster centre, as the system number density per magnitude bin. {We emphasize that the luminosity function} is our primary observational result, independent of evolutionary and atmospheric modelling. In order to compare with other studies, we discuss the properties of the mass function. 

\section{Mass function}  \label{mf}

We transform the luminosity of the new candidates using the mass-luminosity relation provided by \citet{Baraf98}, used in the $K_s$ band, as the models prove to better match the observations for that filter (R11).
For the previously published candidates we use the masses reported in R11.
We also try the new BT-Settl isochrones of \citet{Allar11}. We find only slightly larger masses, with the difference picking at less than 5\% around $0.3\,{\rm M_{\odot}}$.
We assume that the objects are individual stars and do not make correction for binarity,  {because of the lack of specific statistics describing the binarity as a function of spectral type. }

\begin{figure}
\includegraphics[width=.5\textwidth]{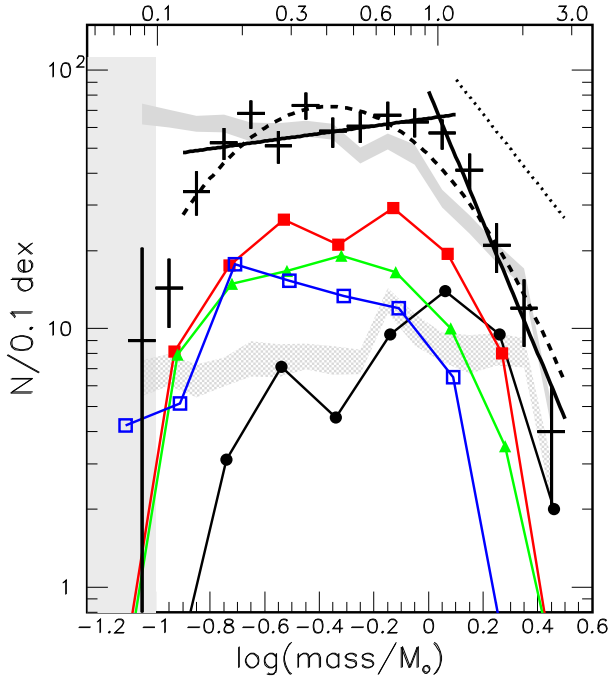}
\caption{Corrected mass function (number of candidates {with $|\rm{v}_\perp|<2\rm\,km\,s^{-1}$}) of the Hyades in the 2MASS $K_s$ band, for regions of various cluster radii: 0 to 3.1 \,pc (circle); 3.1 to 9\,pc (filled square); 9 to 18 \,pc (triangle); and 18 to 30 \,pc (open square). 
The top histogram shows the complete mass function up to 30\,pc, fitted with two power laws (thick lines), and a log-normal mass function (dashed line). The Salpeter mass function is shown for comparison (dotted line).
The grey areas are the 1-$\sigma$ locus of the simulation of \citet{Ernst11} for the cluster centre ($r_c<3$\,pc, bottom) and the whole volume ($r_c<30$\,pc, top).
The light grey area for $m<0.1\,{\rm M_{\odot}}$ indicates the masses for which our results are strongly affected by incompleteness. 
}
\label{MF}
\end{figure}

We show the results in Fig.\,\ref{MF}, again for the same four different cluster regions, and for the complete survey up to 30\,pc.
Here we consider the mass function as the number of systems {between $\log m$ and $\log m+ {\rm d} \log m$}, as a function of the logarithm of their mass \citep{Salpe55}:
\begin{align*}
\xi(\log m)=\frac{{\rm d}N(\log m)}{{\rm d} \log m} \propto m^{-\Gamma}=m^{-(\alpha-1)}
\end{align*}

{For the volume within 30\,pc, the} mass function is well  {(minimum $\chi^2$)} fitted for its high-mass (0.80--$3.2\,{\rm M_{\odot}}$) part by a power law with (logarithmic) slope \mbox{$\Gamma=2.53\pm 0.30$} (or \mbox{$\Gamma=2.21\pm 0.37$} over 0.80--$2.0\,{\rm M_{\odot}}$), steeper than the canonical Salpeter mass function ({in this logarithmic representation} the Salpeter power-law would have a slope of $\Gamma=1.35$, i.e. $\alpha=2.35$). 
{In the intermediate part} of the mass spectrum, 0.13--$1.2\,{\rm M_{\odot}}$, the mass function is nearly flat, with \mbox{$\Gamma=0.15\pm 0.06$}. 
{The error estimates on the slope are consistent with the more sophisticated analysis of \citet{Weisz12}.}

Alternatively we can fit the range of 0.13--$3.2\,{\rm M_{\odot}}$ with a log-normal function \citep{Chabr03}, 
\begin{align*}
\mbox{$\xi(m) = e^{-(\log m - \log m_c)^2/2\sigma^2}$}, 
\end{align*}
with parameters: 
\mbox{$m_c=(0.44\pm 0.03)\rm M_{\odot}$} and \mbox{$\sigma=0.39\pm 0.02$}. 

Finally {the last two mass bins} with good completeness, $m=$0.10--$0.15\,{\rm M_{\odot}}$ show a deviation from the power-law mass function, and a significant deficiency in very-low-mass stars. 

\citet{Ernst11} modelled the R11 Hyades mass function using an $N$-body simulation of the cluster assuming an analytic Milky Way potential and the \citet{Kroup01} initial mass function down to $0.08\,{\rm M_{\odot}}$.
The agreement is good over the whole volume surveyed ($r_c<30$\,pc, see Fig.\,\ref{MF}).
However the {observed knee} at $1\,{\rm M_{\odot}}$ is smooth in the simulation,  {which may be due to unresolved binaries. 
In addition, in the mass bin 0.1--$0.12\,{\rm M_{\odot}}$, we see a clear decrease in the mass function, while the model displays a small increase. 
This conclusion is only based on one mass bin in our own data, but seems consistent with the results of \citet{Bouvi08}. 
Similarly}, for the cluster centre ($r_c<3$\,pc), the simulation predicts a flat mass function down to the end of the main sequence, while we clearly see a continuous decrease for low-mass stars.
{However the simulation by \citet{Ernst11} did not take into account encounters with large molecular clouds, which may strip the cluster of its low-mass members, and this could the reason for the observed discrepancy. 
Alternatively, assuming that a fraction of stars are members of unresolved binaries also leads to a decrease of the (system) mass function at the low-mass end (Ernst, {\it priv. com.}).
}

\begin{figure}
\includegraphics[width=.5\textwidth]{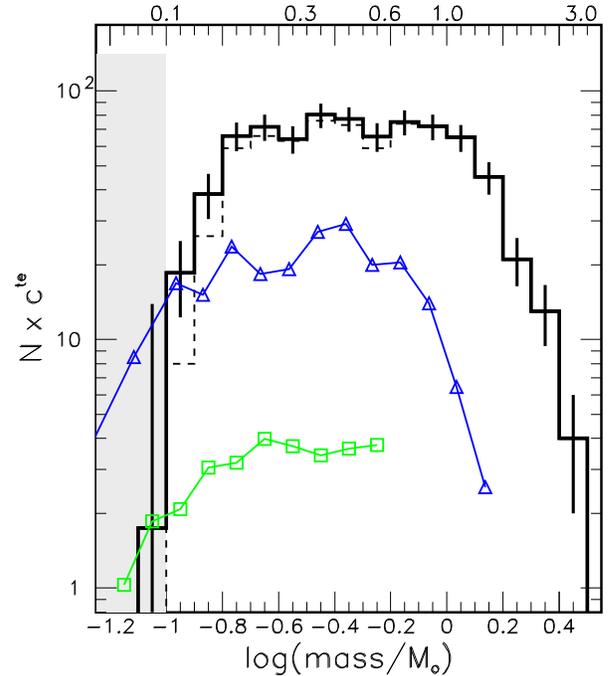}
\caption{Mass functions of the Hyades: this study (thick histogram, {$|\rm{v}_\perp|<4\rm\,km\,s^{-1}$ for consistency}); R11 (thin dashed line); and \citet{Bouvi08} (triangle); and of the Praesepe cluster:  
\citet{Boudr12} (square).
The light grey area for $m<0.1\,{\rm M_{\odot}}$ indicates the masses for which our results are strongly affected by incompleteness. 
}
\label{MFcomp}
\end{figure}

In Fig.\,\ref{MFcomp}, we compare our mass functions with published results of  {another nearby and well observed} intermediate-age cluster, the Praesepe cluster \citep{Boudr12}, as well as previous mass functions of the Hyades \citep[][R11]{Bouvi08}.

Compared to R11, we determine the mass function for lower masses, with the consequence that the mass function plateau is extended towards lower masses, around $0.13\,{\rm M_{\odot}}$.
Our result is similar to the Hyades mass function of \citet{Bouvi08}, but with better statistics thanks to the large volume probed. Therefore insignificant features such as the dip at $0.25\,{\rm M_{\odot}}$ is not seen in our mass function. In addition, our mass function is flatter because of the contribution of the {Hyads} at large cluster radii. 

{Comparing with the mass function of \citet{Boudr12}, the Praesepe cluster seems to have retained its low-mass members at a higher rate, or has a steeper initial mass function, than the Hyades cluster}. We point out that the different surveys use different selection methods and, of course, data sets.
Thus they have different contamination rate and completeness. In that respect, our combined use of proper motion and multi-band photometry for a nearby cluster, makes our study less prone to systematic errors than {purely} photometric studies. 

\section{Spatial structure} \label{spatstruc}

 The total mass of the additional {62}~candidates, within 30\,pc, is $11\,\rm M_\odot$.
By correcting for incompleteness as outlined above, we find about {14}$\,\rm M_\odot$.
Within 9\,pc of the centre, the {18} new candidates make up less than $3\,\rm M_\odot$.

In both cases, this is a small correction to the previous estimate by R11, i.e. $276\,\rm M_\odot$ within 9\,pc. 
Therefore, we do not update the analysis presented in R11 as the results would be very similar.

\begin{figure*}
\includegraphics[width=\textwidth,angle=0]{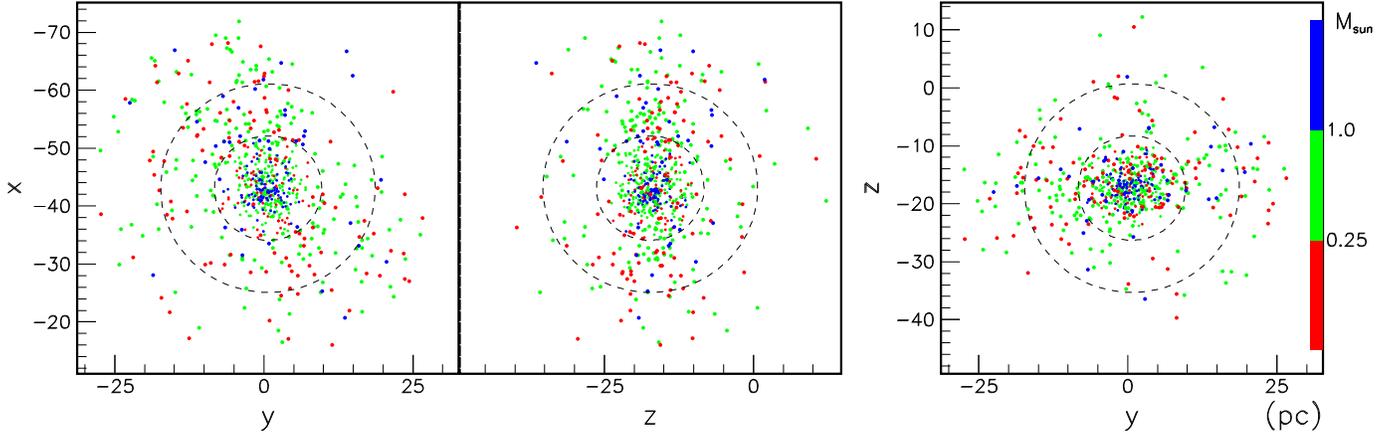}
\caption{Projection of the candidates {with $|\rm{v}_\perp|<2\,\rm km s^{-1}$} on the Galactic coordinate planes.
The colour scales with the mass (inner candidates within 9\,pc of the centre are shown as smaller dots for clarity).
The 9-pc- and 18-pc-radius circles are shown as dashed lines.
}
\label{xyz}
\end{figure*}

In Fig.\,\ref{xyz} we show the distribution of the candidates {with $|\rm{v}_\perp|<2\,\rm km\,s^{-1}$} in cartesian Galactic coordinates, colour-coded by mass. It is clear that massive (blue) stars concentrate on the centre, while low-mass members are distributed over {the whole area}.
The ellipticity of the cluster is clearly visible \citep[Fig.6 of ][]{Ernst11}.

\newcommand {\Mmst}         {\mbox{${\cal{M}}_{\mathrm{MST}}^{\mathrm{\Gamma}}$}}
\newcommand {\gmst}[1]      {\mbox{$\gamma_{\mathrm{MST}}^{#1}$}}
\newcommand {\gmstMean}[1]  {\mbox{$\langle\gmst{#1}\rangle$}}
\newcommand {\gmstDelta}[1] {\mbox{$\Delta\gmst{#1}$}}
\newcommand {\grefMean}     {\gmstMean{\mathrm{ref}}}
\newcommand {\grefDelta}    {\gmstDelta{\mathrm{ref}}}
\newcommand {\gmass}        {\gmst{\mathrm{mass}}}
\newcommand {\Gmst}         {\mbox{$\Gamma_{\mathrm{MST}}$}}
\newcommand {\GmstDelta}    {\mbox{$\Delta\Gmst$}}

As a proxy for mass segregation we use the minimum spanning tree (MST) method $\Mmst$ developed by \citet{Olcza11}. {The MST is the graph connecting
all $n$ vertices within a given sample $\zeta$ with the lowest possible sum of the $(n-1)$ edge lengths $e_{i, \zeta}$ excluding closed loops
\citep{Gower69}. The authors use the geometrical mean to define a measure of the MST of a sample $\zeta$,

\begin{equation}
  \begin{aligned}
    \gmst{\zeta} &= \exp\left[ \frac{1}{n} \sum_{i=1}^n \ln e_{i, \zeta} \right] \,.\\
  \end{aligned}
\end{equation}

with the corresponding statistics for a set of $m$ equal-sized samples $\zeta_j$,

\begin{equation}
  \begin{aligned}
    \gmstMean{\zeta_j}&= \frac{1}{m} \sum_{j=1}^{m} \gmst{\zeta_j} \,,\\
    \gmstDelta{\zeta_j}&= \sqrt{ \frac{1}{m-1} \sum_{j=1}^{m} \left( \gmst{\zeta_j} - \gmstMean{\zeta_j} \right)^2 } \,.
  \end{aligned}
\end{equation}

Based on these quantities we define a normalized measure of mass segregation of a population of $N$ stars,}
\begin{equation}
  \begin{aligned}
    \Gmst &= \frac{\grefMean}{\gmass} \,,\\
    \GmstDelta &= \frac{\grefDelta}{\gmass} \,,
  \end{aligned}
\end{equation}
where the superscript ``ref'' refers to an ensemble of $m$ samples of $n \le N$ random stars, while the superscript ``mass'' refers to a sample of the
$n$ most massive stars. The number $m$ is chosen such that the ensemble of random stars is representative of the entire population.  Hence, $\Gmst$ is
a normalized measure of the concentration of the $n$ most massive stars relative to a representative ensemble of random stars. A value of one marks
the unsegregated state. The larger $\Gmst$ the more concentrated are the massive stars compared to the reference ensemble. {The associated
  standard deviation $\GmstDelta$ quantifies the significance of the result.}

\begin{figure}
\includegraphics[width=1.0\linewidth]{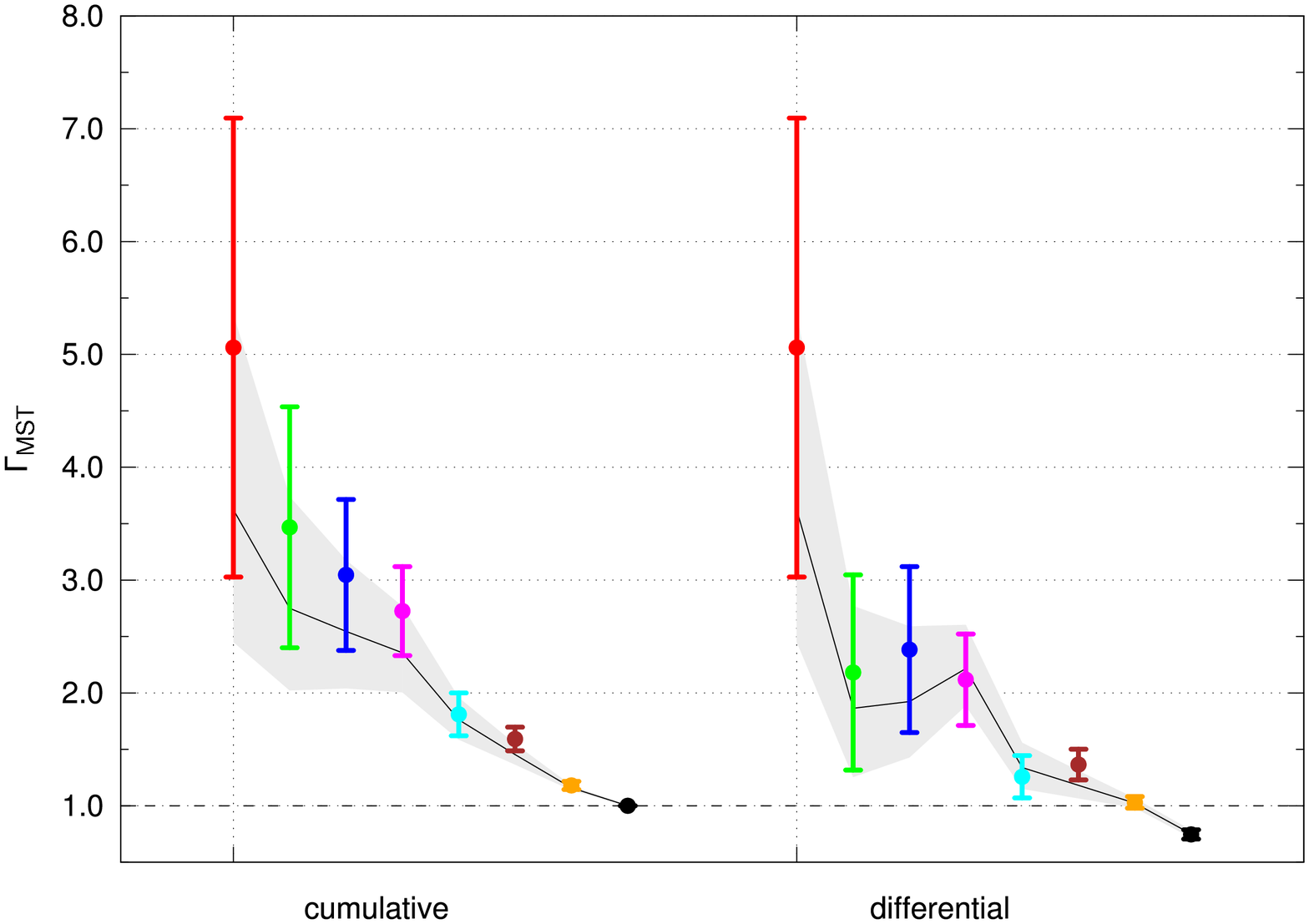}
\includegraphics[width=1.0\linewidth]{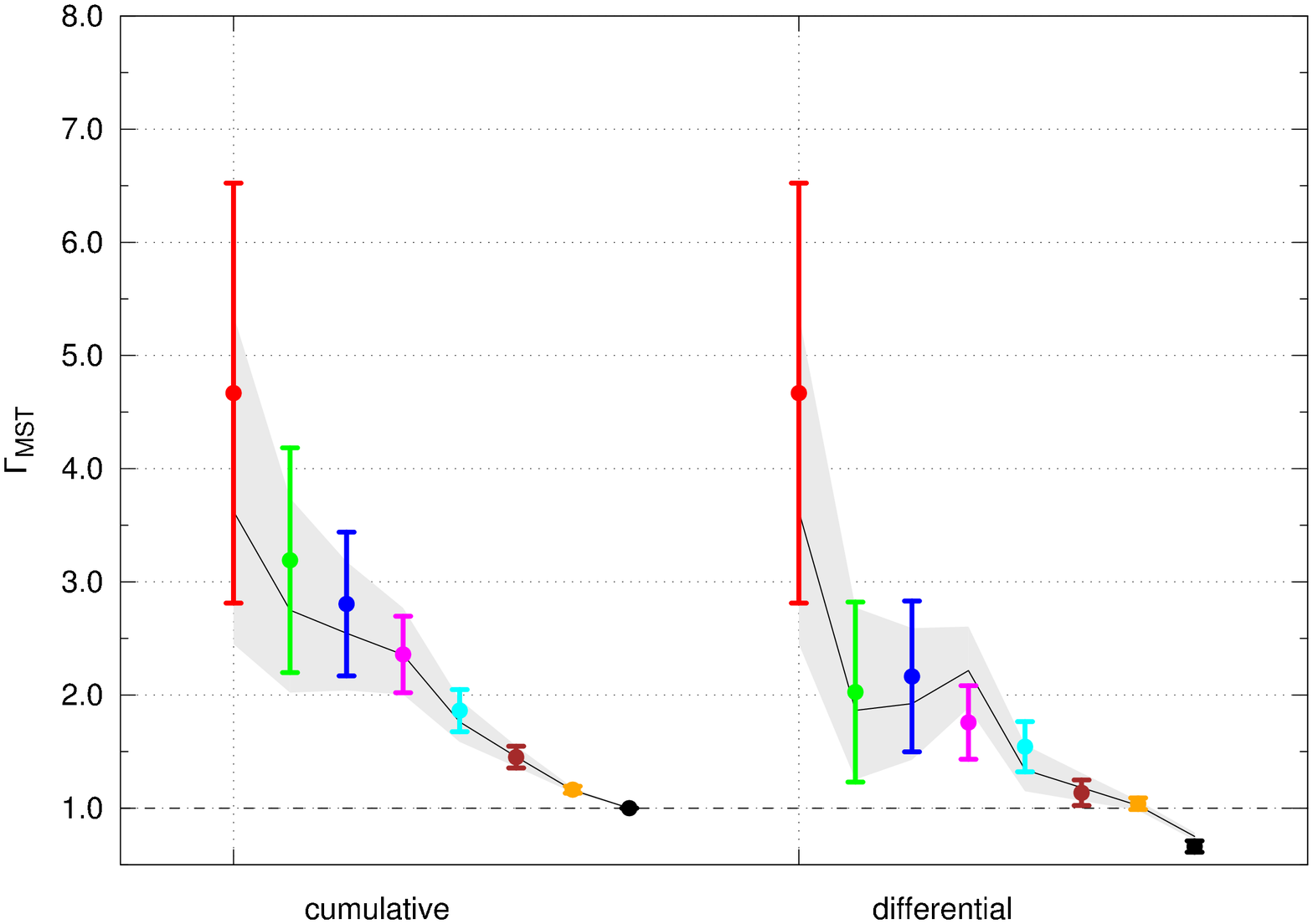}
\caption{Mass segregation analysis of our Hyades data and comparison with \citet{Ernst11}. \emph{Top}: Full sample (candidates with
  $|\rm{v}_\perp|<4\,\rm km s^{-1}$). \emph{Bottom}: Reduced sample with velocity cut as explained in the text and statistical removal of
  contaminants. The left hand-side shows the "cumulative" $\Gamma_{MST}$ (points) for the 5 (red), 10 (green), 20 (blue), 50 (magenta), 100 (cyan),
  200 (brown), 500 (orange) most massive, and all stars (black). The right-hand side shows the ``differential'' $\Gamma_{MST}$ (points) for the 5
  (red), 6--10 (green), 11--20 (blue), 21--50 (magenta), 51--100 (cyan), 101--200 (brown), 201--500 (orange) most massive stars, and the remainder
  (black). The error bars mark the 1--$\sigma$ uncertainties $\Delta\Gamma_{MST}$. The black solid line and the grey region represent $\Gamma_{MST}$
  and $\Delta\Gamma_{MST}$ for the data of R11, respectively.}
\label{segreg}
\end{figure}

{Figure\,\ref{segreg} shows the mass segregation analysis of two sets of our observational sample (points with error bars) compared to the
  observational data of R11 as published in \citet{Ernst11} (black line with shaded error region). The data in the upper panel are based on the full
  sample of 786 stars (with $|\rm{v}_\perp|<4\,\rm km\,s^{-1}$), consisting of 724 candidates from R11 and our 62 new low-mass candidates, mostly
  located in the outer cluster parts. The effect of these new candidates is to increase the normalized mass segregation measure $\Gmst$ of the more
  concentrated subsamples as is most evident for the five most massive stars (red). The result for the reduced sample of the ``best'' 669 candidates
  (with $|\rm{v}_\perp|<2\,\rm km\,s^{-1}$) is presented in the lower panel\footnote{We recall that the massive binary member \object{77 Tau} (PPMXL\,2992334831999166121) is also part of 
    this sample as its large $|\rm{v}_\perp| = 2.3\,\rm km\,s^{-1}$ is caused by the orbital motion.\label{77Tau}}. We have further taken into account
  contamination by field stars in a statistical approach yet found no significant effect. The more conservative membership selection makes the shape
  of the distribution of the differential $\Gmst$ much more regular. This characteristic signature of dynamically evolved mass segregation \citep[see
  e.g.][]{Ernst11} implies a larger purity of the reduced sample and hence justifies the lower velocity threshold.}

Our main findings from the mass segregation analysis of the {revised} Hyades {candidates} are the following:
\begin{enumerate}
\item The cumulative plot shows a significant degree of mass segregation for all samples. Comparison with the differential samples demonstrates the
  exceptionally compact configuration of the five most massive stars (red). The steady decline of $\Gmst$ with stellar mass for the differential
  samples is a characteristic signature of dynamically evolved mass segregation \citep[cf.][]{Ernst11}.
\item The least massive Hyades candidates are clearly showing ``inverse'' mass segregation: $\Gmst < 1$ for the last bin in the differential
  plot. This is another characteristic signature of dynamically evolved mass segregation that has removed the lowest mass members from the inner
  cluster parts.
\item The signature of mass segregation agrees fairly well with a moderately \citep[$S = 0.3$: cf.][]{Subr08} mass segregated star cluster model with
  1000 members as shown in the middle panel of Fig. 4 in \citet{Olcza11}.
\item Our extended Hyades membership sample with {48}~additional low-mass cluster {candidates} compared to the data of R11 demonstrates that
  the Hyades are stronger mass segregated than estimated before \citep{Ernst11}.
\end{enumerate}

\section{Conclusion} \label{discussion}

We have combined the PPMXL and Pan-STARRS1 catalogues to search for low-mass Hyades members up to 30\,pc from the cluster centre.
We select candidates based on the PPMXL kinematics and Pan-STARRS1 astrometry and photometry, {combined with 2MASS, WISE and SDSS photometry,} to produce a clean sample of candidates nearly complete down to $0.1\,{\rm M_{\odot}}$. 

We discover {54}~new candidates, with an average mass of  {0.18}$\,{\rm M_{\odot}}$ and an average distance from the cluster centre of {18}\,pc. 
We select a purer sample of {43}~ {new candidates with velocity perpendicular to the Hyades motion smaller than $2\rm\,km\,s^{-1}$. 
This doubles the number of Hyades candidates for masses smaller than $0.15\,{\rm M_{\odot}}$.}

{The combination of the astrometric selection of the fast Hyades members, relatively to the field stars, and multi-band photometric selection allows to increase the purity of our candidate sample to larger cluster radii, compared to others distant or slow clusters, and compared to studies using more limited data sets.}
{Using the Besan\c{c}on model to produce a fake catalogue of field stars, the estimated field dwarf contamination is below 20\% for the M-type dwarf in the outer shell, with distances between 18 and 30\,pc, and well below 10\% up to 18\,pc {in any mass bin}.
Our results are based on a detailed analysis of the sensitivity and contamination of our survey.}

We find that the mass function is nearly flat for masses between 0.1 and $1\,{\rm M_{\odot}}$, with \mbox{$-\alpha=0.15\pm 0.06$}.
The cluster is clearly segregated in mass, with the lowest-mass members having been removed from the cluster centre.

We identify one known L-type dwarf, 2MASSI~J0230155+270406, as a likely Hyades member. This object, close enough to be accurately characterized, is a benchmark object at the frontier between low-mass stars and brown dwarfs, with a known age, metallicity and distance.

The Hyades cluster is close enough that PS1 will eventually measure the parallaxes of all low-mass star and L-type-dwarf cluster candidates within 30\,pc of the centre. This will provide a strong confirmation for memberships, and allow a detailed analysis of the kinematics, binarity, activity \citep{Seema10} of the cluster members.
In the pre-Gaia area, the Hyades will therefore be a unique test of cluster evolution {and of pre-main sequence evolution of low-mass stars.}
    
\begin{acknowledgements}
We thank Steve Boudreault, Andreas Ernst and France Allard for providing material from their publications, as well as Mario Gennaro and Josh Schlieder for their comments.
\\
This work was supported by Sonderforschungsbereich SFB 881 ``The Milky Way System'' (subproject B6) of the German Research Foundation (DFG).
C.O. appreciates funding by the German Research Foundation (DFG) grant OL 350/1-1, support by NAOC CAS through the Silk Road Project, and  by Global Networks and Mobility Program of the University of Heidelberg (ZUK 49/1 TP14.8 Spurzem).
M.J. acknowledges support by NASA through Hubble Fellowship grant \#HF-51255.01-A awarded by the Space Telescope Science Institute, which is operated by the Association of Universities for Research in Astronomy, Inc., for NASA, under contract NAS~5-26555.
\\
The Pan-STARRS1 Surveys (PS1) have been made possible through contributions of the Institute for Astronomy, the University of Hawaii, the Pan-STARRS Project Office, the Max-Planck Society and its participating institutes, the Max Planck Institute for Astronomy, Heidelberg and the Max Planck Institute for Extraterrestrial Physics, Garching, The Johns Hopkins University, Durham University, the University of Edinburgh, Queen's University Belfast, the Harvard-Smithsonian Center for Astrophysics, the Las Cumbres Observatory Global Telescope Network Incorporated, the National Central University of Taiwan, the Space Telescope Science Institute, the National Aeronautics and Space Administration under Grant No. NNX08AR22G issued through the Planetary Science Division of the NASA Science Mission Directorate, the National Science Foundation under Grant No. AST-1238877, and the University of Maryland.
\\
Funding for SDSS-III has been provided by the Alfred P. Sloan Foundation, the Participating Institutions, the National Science Foundation, and the U.S. Department of Energy Office of Science. The SDSS-III web site is http://www.sdss3.org/.
\\
{This publication makes use of data products from the Wide-field Infrared Survey Explorer, which is a joint project of the University of California, Los Angeles, and the Jet Propulsion Laboratory/California Institute of Technology, funded by the National Aeronautics and Space Administration.}
\\
This research has made use of the SIMBAD database, operated at CDS, Strasbourg, France; and of the Washington Double Star Catalog maintained at the U.S. Naval Observatory. 
\end{acknowledgements}

\bibliographystyle{aa}
\bibliography{mybib.bib}

\begin{thebibliography}{76}
\expandafter\ifx\csname natexlab\endcsname\relax\def\natexlab#1{#1}\fi

\bibitem[{{Aihara} {et~al.}(2011){Aihara}, {Allende Prieto}, {An}, {Anderson},
  {Aubourg}, {Balbinot}, {Beers}, {Berlind}, {Bickerton}, {Bizyaev}, {Blanton},
  {Bochanski}, {Bolton}, {Bovy}, {Brandt}, {Brinkmann}, {Brown}, {Brownstein},
  {Busca}, {Campbell}, {Carr}, {Chen}, {Chiappini}, {Comparat}, {Connolly},
  {Cortes}, {Croft}, {Cuesta}, {da Costa}, {Davenport}, {Dawson}, {Dhital},
  {Ealet}, {Ebelke}, {Edmondson}, {Eisenstein}, {Escoffier}, {Esposito},
  {Evans}, {Fan}, {Femen{\'{\i}}a Castell{\'a}}, {Font-Ribera}, {Frinchaboy},
  {Ge}, {Gillespie}, {Gilmore}, {Gonz{\'a}lez Hern{\'a}ndez}, {Gott}, {Gould},
  {Grebel}, {Gunn}, {Hamilton}, {Harding}, {Harris}, {Hawley}, {Hearty}, {Ho},
  {Hogg}, {Holtzman}, {Honscheid}, {Inada}, {Ivans}, {Jiang}, {Johnson},
  {Jordan}, {Jordan}, {Kazin}, {Kirkby}, {Klaene}, {Knapp}, {Kneib},
  {Kochanek}, {Koesterke}, {Kollmeier}, {Kron}, {Lampeitl}, {Lang}, {Le Goff},
  {Lee}, {Lin}, {Long}, {Loomis}, {Lucatello}, {Lundgren}, {Lupton}, {Ma},
  {MacDonald}, {Mahadevan}, {Maia}, {Makler}, {Malanushenko}, {Malanushenko},
  {Mandelbaum}, {Maraston}, {Margala}, {Masters}, {McBride}, {McGehee},
  {McGreer}, {M{\'e}nard}, {Miralda-Escud{\'e}}, {Morrison}, {Mullally},
  {Muna}, {Munn}, {Murayama}, {Myers}, {Naugle}, {Fausti Neto}, {Cuong Nguyen},
  {Nichol}, {O'Connell}, {Ogando}, {Olmstead}, {Oravetz}, {Padmanabhan},
  {Palanque-Delabrouille}, {Pan}, {Pandey}, {P{\^a}ris}, {Percival},
  {Petitjean}, {Pfaffenberger}, {Pforr}, {Phleps}, {Pichon}, {Pieri}, {Prada},
  {Price-Whelan}, {Raddick}, {Ramos}, {Reyl{\'e}}, {Rich}, {Richards}, {Rix},
  {Robin}, {Rocha-Pinto}, {Rockosi}, {Roe}, {Rollinde}, {Ross}, {Ross},
  {Rossetto}, {S{\'a}nchez}, {Sayres}, {Schlegel}, {Schlesinger}, {Schmidt},
  {Schneider}, {Sheldon}, {Shu}, {Simmerer}, {Simmons}, {Sivarani}, {Snedden},
  {Sobeck}, {Steinmetz}, {Strauss}, {Szalay}, {Tanaka}, {Thakar}, {Thomas},
  {Tinker}, {Tofflemire}, {Tojeiro}, {Tremonti}, {Vandenberg}, {Vargas
  Maga{\~n}a}, {Verde}, {Vogt}, {Wake}, {Wang}, {Weaver}, {Weinberg}, {White},
  {White}, {Yanny}, {Yasuda}, {Yeche}, \& {Zehavi}}]{SDSS_DR8}
{Aihara}, H., {Allende Prieto}, C., {An}, D., {et~al.} 2011, \apjs, 193, 29

\bibitem[{{Allard} {et~al.}(2010){Allard}, {Homeier}, \& {Freytag}}]{Allar11}
{Allard}, F., {Homeier}, D., \& {Freytag}, B. 2010, ArXiv {\tt 1011.5405}

\bibitem[{{Andrei} {et~al.}(2011){Andrei}, {Smart}, {Penna}, {d'Avila},
  {Bucciarelli}, {Camargo}, {Crosta}, {Dapr{\`a}}, {Goldman}, {Jones},
  {Lattanzi}, {Nicastro}, {Pinfield}, {da Silva Neto}, \& {Teixeira}}]{Andre11}
{Andrei}, A.~H., {Smart}, R.~L., {Penna}, J.~L., {et~al.} 2011, \aj, 141, 54

\bibitem[{{Baraffe} {et~al.}(1998){Baraffe}, {Chabrier}, {Allard}, \&
  {Hauschildt}}]{Baraf98}
{Baraffe}, I., {Chabrier}, G., {Allard}, F., \& {Hauschildt}, P.~H. 1998, \aap,
  337, 403

\bibitem[{{Beaumont} \& {Magnier}(2010)}]{Beaum10}
{Beaumont}, C.~N. \& {Magnier}, E.~A. 2010, \pasp, 122, 1389

\bibitem[{{Berger} {et~al.}(2010){Berger}, {Basri}, {Fleming}, {Giampapa},
  {Gizis}, {Liebert}, {Mart{\'{\i}}n}, {Phan-Bao}, \& {Rutledge}}]{Berge10}
{Berger}, E., {Basri}, G., {Fleming}, T.~A., {et~al.} 2010, \apj, 709, 332

\bibitem[{{Blake} {et~al.}(2010){Blake}, {Charbonneau}, \& {White}}]{Blake10}
{Blake}, C.~H., {Charbonneau}, D., \& {White}, R.~J. 2010, \apj, 723, 684

\bibitem[{{Bochanski} {et~al.}(2005){Bochanski}, {Hawley}, {Reid}, {Covey},
  {West}, {Tinney}, \& {Gizis}}]{Bocha05}
{Bochanski}, J.~J., {Hawley}, S.~L., {Reid}, I.~N., {et~al.} 2005, \aj, 130,
  1871

\bibitem[{{Boudreault} {et~al.}(2012){Boudreault}, {Lodieu}, {Deacon}, \&
  {Hambly}}]{Boudr12}
{Boudreault}, S., {Lodieu}, N., {Deacon}, N.~R., \& {Hambly}, N.~C. 2012,
  \mnras, 426, 3419

\bibitem[{{Bouvier} {et~al.}(2008){Bouvier}, {Kendall}, {Meeus}, {Testi},
  {Moraux}, {Stauffer}, {James}, {Cuillandre}, {Irwin}, {McCaughrean},
  {Baraffe}, \& {Bertin}}]{Bouvi08}
{Bouvier}, J., {Kendall}, T., {Meeus}, G., {et~al.} 2008, \aap, 481, 661

\bibitem[{{Bryja} {et~al.}(1994){Bryja}, {Humphreys}, \& {Jones}}]{Bryja94}
{Bryja}, C., {Humphreys}, R.~M., \& {Jones}, T.~J. 1994, \aj, 107, 246

\bibitem[{{Bryja} {et~al.}(1992){Bryja}, {Jones}, {Humphreys}, {Lawrence},
  {Pennington}, \& {Zumach}}]{Bryja92}
{Bryja}, C., {Jones}, T.~J., {Humphreys}, R.~M., {et~al.} 1992, \apjl, 388, L23

\bibitem[{{Chabrier}(2003)}]{Chabr03}
{Chabrier}, G. 2003, \pasp, 115, 763

\bibitem[{{Close} {et~al.}(2003){Close}, {Siegler}, {Freed}, \&
  {Biller}}]{Close03}
{Close}, L.~M., {Siegler}, N., {Freed}, M., \& {Biller}, B. 2003, \apj, 587,
  407

\bibitem[{{Copenhagen University Obs.}(2006)}]{Copen06}
{Copenhagen University Obs.}, {Institute of Astronomy, Cambridge, UK}, R., ed.
  2006, {Carlsberg Meridian Catalog Number 14}

\bibitem[{{Cruz} \& {Reid}(2002)}]{Cruz02}
{Cruz}, K.~L. \& {Reid}, I.~N. 2002, \aj, 123, 2828

\bibitem[{{Cruz} {et~al.}(2007){Cruz}, {Reid}, {Kirkpatrick}, {Burgasser},
  {Liebert}, {Solomon}, {Schmidt}, {Allen}, {Hawley}, \& {Covey}}]{Cruz07}
{Cruz}, K.~L., {Reid}, I.~N., {Kirkpatrick}, J.~D., {et~al.} 2007, \aj, 133,
  439

\bibitem[{{Cruz} {et~al.}(2003){Cruz}, {Reid}, {Liebert}, {Kirkpatrick}, \&
  {Lowrance}}]{Cruz03}
{Cruz}, K.~L., {Reid}, I.~N., {Liebert}, J., {Kirkpatrick}, J.~D., \&
  {Lowrance}, P.~J. 2003, \aj, 126, 2421

\bibitem[{{Deacon} {et~al.}(2009){Deacon}, {Hambly}, {King}, \&
  {McCaughrean}}]{Deaco09}
{Deacon}, N.~R., {Hambly}, N.~C., {King}, R.~R., \& {McCaughrean}, M.~J. 2009,
  \mnras, 394, 857

\bibitem[{{DeGennaro} {et~al.}(2009){DeGennaro}, {von Hippel}, {Jefferys},
  {Stein}, {van Dyk}, \& {Jeffery}}]{DeGen09}
{DeGennaro}, S., {von Hippel}, T., {Jefferys}, W.~H., {et~al.} 2009, \apj, 696,
  12

\bibitem[{{Delorme} {et~al.}(2011){Delorme}, {Collier Cameron}, {Hebb},
  {Rostron}, {Lister}, {Norton}, {Pollacco}, \& {West}}]{Delor11}
{Delorme}, P., {Collier Cameron}, A., {Hebb}, L., {et~al.} 2011, \mnras, 413,
  2218

\bibitem[{{Dupuy} \& {Liu}(2011)}]{Dupuy11}
{Dupuy}, T.~J. \& {Liu}, M.~C. 2011, \apj, 733, 122

\bibitem[{{Ernst} {et~al.}(2011){Ernst}, {Just}, {Berczik}, \&
  {Olczak}}]{Ernst11}
{Ernst}, A., {Just}, A., {Berczik}, P., \& {Olczak}, C. 2011, \aap, 536, A64

\bibitem[{{Faherty} {et~al.}(2009){Faherty}, {Burgasser}, {Cruz}, {Shara},
  {Walter}, \& {Gelino}}]{Faher09}
{Faherty}, J.~K., {Burgasser}, A.~J., {Cruz}, K.~L., {et~al.} 2009, \aj, 137, 1

\bibitem[{{Frankowski} {et~al.}(2007){Frankowski}, {Jancart}, \&
  {Jorissen}}]{Frank07}
{Frankowski}, A., {Jancart}, S., \& {Jorissen}, A. 2007, \aap, 464, 377

\bibitem[{{G{\'a}lvez-Ortiz} {et~al.}(2010){G{\'a}lvez-Ortiz}, {Clarke},
  {Pinfield}, {Jenkins}, {Folkes}, {P{\'e}rez}, {Day-Jones}, {Burningham},
  {Jones}, {Barnes}, \& {Pokorny}}]{Galvez10}
{G{\'a}lvez-Ortiz}, M.~C., {Clarke}, J.~R.~A., {Pinfield}, D.~J., {et~al.}
  2010, \mnras, 409, 552

\bibitem[{{Giclas} {et~al.}(1962){Giclas}, {Burnham}, \& {Thomas}}]{Gicla62}
{Giclas}, H.~L., {Burnham}, R., \& {Thomas}, N.~G. 1962, Lowell Observatory
  Bulletin, 5, 257

\bibitem[{{Gizis} {et~al.}(2000){Gizis}, {Monet}, {Reid}, {Kirkpatrick},
  {Liebert}, \& {Williams}}]{Gizis00}
{Gizis}, J.~E., {Monet}, D.~G., {Reid}, I.~N., {et~al.} 2000, \aj, 120, 1085

\bibitem[{Gower \& Ross(1969)}]{Gower69}
Gower, J.~C. \& Ross, G. J.~S. 1969, Journal of the Royal Statistical Society.
  Series C (Applied Statistics), 18, pp. 54

\bibitem[{{Hambly} {et~al.}(2008){Hambly}, {Collins}, {Cross}, {Mann}, {Read},
  {Sutorius}, {Bond}, {Bryant}, {Emerson}, {Lawrence}, {Rimoldini}, {Stewart},
  {Williams}, {Adamson}, {Hirst}, {Dye}, \& {Warren}}]{Hambl08}
{Hambly}, N.~C., {Collins}, R.~S., {Cross}, N.~J.~G., {et~al.} 2008, \mnras,
  384, 637

\bibitem[{{Hogan} {et~al.}(2008){Hogan}, {Jameson}, {Casewell}, {Osbourne}, \&
  {Hambly}}]{Hogan08}
{Hogan}, E., {Jameson}, R.~F., {Casewell}, S.~L., {Osbourne}, S.~L., \&
  {Hambly}, N.~C. 2008, \mnras, 388, 495

\bibitem[{{Jordi} {et~al.}(2006){Jordi}, {Grebel}, \& {Ammon}}]{Jordi06}
{Jordi}, K., {Grebel}, E.~K., \& {Ammon}, K. 2006, \aap, 460, 339

\bibitem[{{Kaiser} {et~al.}(2002){Kaiser}, {Aussel}, {Burke}, {Boesgaard},
  {Chambers}, {Chun}, {Heasley}, {Hodapp}, {Hunt}, {Jedicke}, {Jewitt},
  {Kudritzki}, {Luppino}, {Maberry}, {Magnier}, {Monet}, {Onaka}, {Pickles},
  {Rhoads}, {Simon}, {Szalay}, {Szapudi}, {Tholen}, {Tonry}, {Waterson}, \&
  {Wick}}]{Kaise02}
{Kaiser}, N., {Aussel}, H., {Burke}, B.~E., {et~al.} 2002, in Society of
  Photo-Optical Instrumentation Engineers (SPIE) Conference Series, Vol. 4836,
  Society of Photo-Optical Instrumentation Engineers (SPIE) Conference Series,
  ed. {J.~A.~Tyson \& S.~Wolff}, 154--164

\bibitem[{{Kharchenko} \& {Schilbach}(1996)}]{Kharc96}
{Kharchenko}, N. \& {Schilbach}, E. 1996, Baltic Astronomy, 5, 337

\bibitem[{{Kirkpatrick} {et~al.}(1997){Kirkpatrick}, {Beichman}, \&
  {Skrutskie}}]{Kirkp97}
{Kirkpatrick}, J.~D., {Beichman}, C.~A., \& {Skrutskie}, M.~F. 1997, \apj, 476,
  311

\bibitem[{{Kroupa}(2001)}]{Kroup01}
{Kroupa}, P. 2001, \mnras, 322, 231

\bibitem[{{Lawrence} {et~al.}(2007){Lawrence}, {Warren}, {Almaini}, {Edge},
  {Hambly}, {Jameson}, {Lucas}, {Casali}, {Adamson}, {Dye}, {Emerson},
  {Foucaud}, {Hewett}, {Hirst}, {Hodgkin}, {Irwin}, {Lodieu}, {McMahon},
  {Simpson}, {Smail}, {Mortlock}, \& {Folger}}]{Lawre07}
{Lawrence}, A., {Warren}, S.~J., {Almaini}, O., {et~al.} 2007, \mnras, 379,
  1599

\bibitem[{{Luhman}(2006)}]{Luhma06sd}
{Luhman}, K.~L. 2006, \apj, 645, 676

\bibitem[{{Magnier}(2006)}]{Magni06}
{Magnier}, E. 2006, in The Advanced Maui Optical and Space Surveillance
  Technologies Conference

\bibitem[{{Magnier} {et~al.}(2008){Magnier}, {Liu}, {Monet}, \&
  {Chambers}}]{Magni08}
{Magnier}, E.~A., {Liu}, M., {Monet}, D.~G., \& {Chambers}, K.~C. 2008, in IAU
  Symposium, Vol. 248, IAU Symposium, ed. {W.~J.~Jin, I.~Platais, \&
  M.~A.~C.~Perryman}, 553--559

\bibitem[{{McArthur} {et~al.}(2011){McArthur}, {Benedict}, {Harrison}, \& {van
  Altena}}]{McArt11}
{McArthur}, B.~E., {Benedict}, G.~F., {Harrison}, T.~E., \& {van Altena}, W.
  2011, \aj, 141, 172

\bibitem[{{Olczak} {et~al.}(2011){Olczak}, {Spurzem}, \& {Henning}}]{Olcza11}
{Olczak}, C., {Spurzem}, R., \& {Henning}, T. 2011, \aap, 532, 119

\bibitem[{{Padmanabhan} {et~al.}(2008){Padmanabhan}, {Schlegel}, {Finkbeiner},
  {Barentine}, {Blanton}, {Brewington}, {Gunn}, {Harvanek}, {Hogg},
  {Ivezi{\'c}}, {Johnston}, {Kent}, {Kleinman}, {Knapp}, {Krzesinski}, {Long},
  {Neilsen}, {Nitta}, {Loomis}, {Lupton}, {Roweis}, {Snedden}, {Strauss}, \&
  {Tucker}}]{Padma08}
{Padmanabhan}, N., {Schlegel}, D.~J., {Finkbeiner}, D.~P., {et~al.} 2008, \apj,
  674, 1217

\bibitem[{{Perryman} {et~al.}(1998){Perryman}, {Brown}, {Lebreton}, {Gomez},
  {Turon}, {Cayrel de Strobel}, {Mermilliod}, {Robichon}, {Kovalevsky}, \&
  {Crifo}}]{Perry98}
{Perryman}, M.~A.~C., {Brown}, A.~G.~A., {Lebreton}, Y., {et~al.} 1998, \aap,
  331, 81

\bibitem[{{Phan-Bao} \& {Bessell}(2006)}]{PhanB06}
{Phan-Bao}, N. \& {Bessell}, M.~S. 2006, \aap, 446, 515

\bibitem[{{Reid} {et~al.}(2008{\natexlab{a}}){Reid}, {Cruz}, {Burgasser}, \&
  {Liu}}]{Reid08}
{Reid}, I.~N., {Cruz}, K.~L., {Burgasser}, A.~J., \& {Liu}, M.~C.
  2008{\natexlab{a}}, \aj, 135, 580

\bibitem[{{Reid} {et~al.}(2008{\natexlab{b}}){Reid}, {Cruz}, {Kirkpatrick},
  {Allen}, {Mungall}, {Liebert}, {Lowrance}, \& {Sweet}}]{Reid08X}
{Reid}, I.~N., {Cruz}, K.~L., {Kirkpatrick}, J.~D., {et~al.}
  2008{\natexlab{b}}, \aj, 136, 1290

\bibitem[{{Reid} {et~al.}(2002){Reid}, {Kirkpatrick}, {Liebert}, {Gizis},
  {Dahn}, \& {Monet}}]{Reid02h}
{Reid}, I.~N., {Kirkpatrick}, J.~D., {Liebert}, J., {et~al.} 2002, \aj, 124,
  519

\bibitem[{{Reid} \& {Mahoney}(2000)}]{Reid00h}
{Reid}, I.~N. \& {Mahoney}, S. 2000, \mnras, 316, 827

\bibitem[{{Reid}(1992)}]{Reid92}
{Reid}, N. 1992, \mnras, 257, 257

\bibitem[{{Reid}(1993)}]{Reid93}
{Reid}, N. 1993, \mnras, 265, 785

\bibitem[{{Reiners} \& {Basri}(2008)}]{Reine08}
{Reiners}, A. \& {Basri}, G. 2008, \apj, 684, 1390

\bibitem[{{Robin} {et~al.}(2003){Robin}, {Reyl{\'e}}, {Derri{\`e}re}, \&
  {Picaud}}]{Robin03}
{Robin}, A.~C., {Reyl{\'e}}, C., {Derri{\`e}re}, S., \& {Picaud}, S. 2003,
  \aap, 409, 523

\bibitem[{{R\"{o}ser} {et~al.}(2010){R\"{o}ser}, {Demleitner}, \&
  {Schilbach}}]{Roese10}
{R\"{o}ser}, S., {Demleitner}, M., \& {Schilbach}, E. 2010, \aj, 139, 2440

\bibitem[{{R{\"o}ser} {et~al.}(2011){R{\"o}ser}, {Schilbach}, {Piskunov},
  {Kharchenko}, \& {Scholz}}]{Roese11}
{R{\"o}ser}, S., {Schilbach}, E., {Piskunov}, A.~E., {Kharchenko}, N.~V., \&
  {Scholz}, R.-D. 2011, \aap, 531, 92

\bibitem[{{Salpeter}(1955)}]{Salpe55}
{Salpeter}, E.~E. 1955, \apj, 121, 161

\bibitem[{{Schlafly} {et~al.}(2012){Schlafly}, {Finkbeiner}, {Juri{\'c}},
  {Magnier}, {Burgett}, {Chambers}, {Grav}, {Hodapp}, {Kaiser}, {Kudritzki},
  {Martin}, {Morgan}, {Price}, {Rix}, {Stubbs}, {Tonry}, \&
  {Wainscoat}}]{Schla12}
{Schlafly}, E.~F., {Finkbeiner}, D.~P., {Juri{\'c}}, M., {et~al.} 2012, \apj,
  756, 158

\bibitem[{{Schlegel} {et~al.}(1998){Schlegel}, {Finkbeiner}, \&
  {Davis}}]{Schle98}
{Schlegel}, D.~J., {Finkbeiner}, D.~P., \& {Davis}, M. 1998, \apj, 500, 525

\bibitem[{{Seemann} {et~al.}(2010){Seemann}, {Reiners}, {Seifahrt}, \&
  {K{\"u}rster}}]{Seema10}
{Seemann}, U., {Reiners}, A., {Seifahrt}, A., \& {K{\"u}rster}, M. 2010, ArXiv
  {\tt 1012.1817}

\bibitem[{{Seifahrt} {et~al.}(2010){Seifahrt}, {Reiners}, {Almaghrbi}, \&
  {Basri}}]{Seifa10}
{Seifahrt}, A., {Reiners}, A., {Almaghrbi}, K.~A.~M., \& {Basri}, G. 2010,
  \aap, 512, 37

\bibitem[{{Siegler} {et~al.}(2003){Siegler}, {Close}, {Mamajek}, \&
  {Freed}}]{Siegl03}
{Siegler}, N., {Close}, L.~M., {Mamajek}, E.~E., \& {Freed}, M. 2003, \apj,
  598, 1265

\bibitem[{{Slesnick} {et~al.}(2006){Slesnick}, {Carpenter}, \&
  {Hillenbrand}}]{Slesn06}
{Slesnick}, C.~L., {Carpenter}, J.~M., \& {Hillenbrand}, L.~A. 2006, \aj, 131,
  3016

\bibitem[{{Stubbs} {et~al.}(2010){Stubbs}, {Doherty}, {Cramer}, {Narayan},
  {Brown}, {Lykke}, {Woodward}, \& {Tonry}}]{Stubb10}
{Stubbs}, C.~W., {Doherty}, P., {Cramer}, C., {et~al.} 2010, \apjs, 191, 376

\bibitem[{{Tamburini} {et~al.}(2002){Tamburini}, {Ortolani}, \&
  {Bianchini}}]{Tambu02}
{Tamburini}, F., {Ortolani}, S., \& {Bianchini}, A. 2002, \aap, 394, 675

\bibitem[{{Tinney}(1993)}]{Tinne93}
{Tinney}, C.~G. 1993, \apj, 414, 279

\bibitem[{{Tonry} {et~al.}(2012){Tonry}, {Stubbs}, {Lykke}, {Doherty},
  {Shivvers}, {Burgett}, {Chambers}, {Hodapp}, {Kaiser}, {Kudritzki},
  {Magnier}, {Morgan}, {Price}, \& {Wainscoat}}]{Tonry12}
{Tonry}, J.~L., {Stubbs}, C.~W., {Lykke}, K.~R., {et~al.} 2012, \apj, 750, 99

\bibitem[{{{\v S}ubr} {et~al.}(2008){{\v S}ubr}, {Kroupa}, \&
  {Baumgardt}}]{Subr08}
{{\v S}ubr}, L., {Kroupa}, P., \& {Baumgardt}, H. 2008, \mnras, 385, 1673

\bibitem[{{van Leeuwen}(2009)}]{vLeeu09}
{van Leeuwen}, F. 2009, \aap, 497, 209

\bibitem[{{Wang} {et~al.}(2011){Wang}, {Boudreault}, {Goldman}, {Henning},
  {Caballero}, \& {Bailer-Jones}}]{Wang11}
{Wang}, W., {Boudreault}, S., {Goldman}, B., {et~al.} 2011, \aap, 531, A164

\bibitem[{{Watson} {et~al.}(2009){Watson}, {Schr{\"o}der}, {Fyfe}, {Page},
  {Lamer}, {Mateos}, {Pye}, {Sakano}, {Rosen}, {Ballet}, {Barcons}, {Barret},
  {Boller}, {Brunner}, {Brusa}, {Caccianiga}, {Carrera}, {Ceballos}, {Della
  Ceca}, {Denby}, {Denkinson}, {Dupuy}, {Farrell}, {Fraschetti}, {Freyberg},
  {Guillout}, {Hambaryan}, {Maccacaro}, {Mathiesen}, {McMahon}, {Michel},
  {Motch}, {Osborne}, {Page}, {Pakull}, {Pietsch}, {Saxton}, {Schwope},
  {Severgnini}, {Simpson}, {Sironi}, {Stewart}, {Stewart}, {Stobbart}, {Tedds},
  {Warwick}, {Webb}, {West}, {Worrall}, \& {Yuan}}]{Watso09}
{Watson}, M.~G., {Schr{\"o}der}, A.~C., {Fyfe}, D., {et~al.} 2009, \aap, 493,
  339

\bibitem[{{Weisz} {et~al.}(2013){Weisz}, {Fouesneau}, {Hogg}, {Rix}, {Dolphin},
  {Dalcanton}, {Foreman-Mackey}, {Lang}, {Johnson}, {Beerman}, {Bell},
  {Gordon}, {Gouliermis}, {Kalirai}, {Skillman}, \& {Williams}}]{Weisz12}
{Weisz}, D.~R., {Fouesneau}, M., {Hogg}, D.~W., {et~al.} 2013, \apj, 762, 123

\bibitem[{{West} {et~al.}(2008){West}, {Hawley}, {Bochanski}, {Covey}, {Reid},
  {Dhital}, {Hilton}, \& {Masuda}}]{West08}
{West}, A.~A., {Hawley}, S.~L., {Bochanski}, J.~J., {et~al.} 2008, \aj, 135,
  785

\bibitem[{{West} {et~al.}(2005){West}, {Walkowicz}, \& {Hawley}}]{West05}
{West}, A.~A., {Walkowicz}, L.~M., \& {Hawley}, S.~L. 2005, \pasp, 117, 706

\bibitem[{{Wielen} {et~al.}(1999){Wielen}, {Dettbarn}, {Jahrei{\ss}},
  {Lenhardt}, \& {Schwan}}]{Wiele99}
{Wielen}, R., {Dettbarn}, C., {Jahrei{\ss}}, H., {Lenhardt}, H., \& {Schwan},
  H. 1999, \aap, 346, 675

\bibitem[{{Wilson} {et~al.}(2003){Wilson}, {Miller}, {Gizis}, {Skrutskie},
  {Houck}, {Kirkpatrick}, {Burgasser}, \& {Monet}}]{Wilso03}
{Wilson}, J.~C., {Miller}, N.~A., {Gizis}, J.~E., {et~al.} 2003, in IAU
  Symposium, Vol. 211, Brown Dwarfs, ed. {E.~Mart{\'{\i}}n}, 197

\bibitem[{{Wright} {et~al.}(2010){Wright}, {Eisenhardt}, {Mainzer}, {Ressler},
  {Cutri}, {Jarrett}, {Kirkpatrick}, {Padgett}, {McMillan}, {Skrutskie},
  {Stanford}, {Cohen}, {Walker}, {Mather}, {Leisawitz}, {Gautier}, {McLean},
  {Benford}, {Lonsdale}, {Blain}, {Mendez}, {Irace}, {Duval}, {Liu}, {Royer},
  {Heinrichsen}, {Howard}, {Shannon}, {Kendall}, {Walsh}, {Larsen}, {Cardon},
  {Schick}, {Schwalm}, {Abid}, {Fabinsky}, {Naes}, \& {Tsai}}]{Wrigh10}
{Wright}, E.~L., {Eisenhardt}, P.~R.~M., {Mainzer}, A.~K., {et~al.} 2010, \aj,
  140, 1868

\end{thebibliography}

\begin{center}
\begin{landscape}
\longtab{5}{
\begin{longtable}{r@{.}l@{ }r@{.}lr@{$\pm$}lr@{$\pm$}lr@{$\pm$}lr@{$\pm$}lr@{$\pm$}lr@{$\pm$}lr@{$\pm$}lcccc}
  \caption{The {62} Hyades candidates {not included in R11}. J2000.0 coordinates from PPMXL {(rounded)}. Photometry from PS1 
  and 2MASS. $r_c$ is the distance from the cluster centre.
  References are given for those stars described in the literature. 
  } \\
  \hline\hline
   \multicolumn{2}{c}{RA} &  \multicolumn{2}{c}{Dec}  & \multicolumn{2}{c}{ $\mu_\alpha.\cos(\delta)$} &  \multicolumn{2}{c}{$\mu_\delta$}& \multicolumn{2}{c}{kin. $\pi$} & \multicolumn{2}{c}{ $g_{\rm P1}$ or $g$}& \multicolumn{2}{c}{$r_{\rm P1}$ or $r$} & \multicolumn{2}{c}{$i_{\rm P1}$ or $i$} & \multicolumn{2}{c}{$K_s$} & $\rm{v}_\perp$ & $r_c$ & mass & Ref. \\
    \multicolumn{4}{c}{(deg, J2000.0)}  & \multicolumn{2}{c}{(mas\,yr$^{-1}$)} & \multicolumn{2}{c}{(mas\,yr$^{-1}$)} & \multicolumn{2}{c}{(mas)} & \multicolumn{2}{c}{(mag)}& \multicolumn{2}{c}{(mag)} & \multicolumn{2}{c}{(mag)} &  \multicolumn{2}{c}{(mag)} & ($\rm km\,s^{-1}$) & (pc) & ($\rm M_\odot$) \\
   \hline
   \endfirsthead
   \caption{continued.}\\
    \hline\hline
   \multicolumn{2}{c}{RA} &  \multicolumn{2}{c}{Dec}  & \multicolumn{2}{c}{ $\mu_\alpha.\cos(\delta)$} &  \multicolumn{2}{c}{$\mu_\delta$}& \multicolumn{2}{c}{ $\pi$} & \multicolumn{2}{c}{ $g_{\rm P1}$ or $g$}& \multicolumn{2}{c}{$r_{\rm P1}$ or $r$} & \multicolumn{2}{c}{$i_{\rm P1}$ or $i$} & \multicolumn{2}{c}{$K_s$} & $\rm{v}_\perp$ & $r_c$ & mass & Ref.  \\
\hline
\endhead
\hline
\endfoot
  28 & 4133 & +32 & 3229 & +214 &  4 & -24 &  4 & 24.1 & 0.4 & 19.05 & 0.01 & 17.67 & 0.01 & \multicolumn{2}{c}{---} & 11.84 & 0.02 & -0.70 & 29.1 & 0.131 &  \\ 
38 & 6231 & +36 & 2197 & +224 &  4 & -47 &  4 & 27.2 & 0.5 & 18.86 & 0.01 & 17.52 & 0.01 & \multicolumn{2}{c}{---} & 11.47 & 0.02 & +1.31 & 24.6 & 0.138 & 1 \\ 
38 & 9566 & $-$7 & 1893 & +284 &  5 & +82 &  5 & 35.4 & 0.6 & 19.48 & 0.01 & 18.03 & 0.01 & 15.68 & 0.01 & 11.43 & 0.03 & +3.00 & 29.1 & 0.111 & 2,3,4 \\ 
41 & 5522 & +5 & 9406 & +146 &  4 &  +5 &  4 & 18.4 & 0.5 & 17.52 & 0.01 & 16.25 & 0.00 & 14.56 & 0.01 & 11.14 & 0.02 & -1.23 & 25.0 & 0.242 &  \\ 
44 & 8713 & +8 & 4708 & +164 &  5 & +13 &  5 & 21.7 & 0.7 & 19.73 & 0.01 & 18.37 & 0.01 & 16.09 & 0.01 & 12.02 & 0.02 & +1.66 & 18.5 & 0.134 &  \\ 
46 & 0086 & +0 & 7642 & +249 &  4 & +32 &  4 & 33.3 & 0.5 & 18.29 & 0.01 & 16.91 & 0.01 & 14.82 & 0.01 & 10.88 & 0.02 & -0.27 & 23.5 & 0.146 & 5,6,7 \\ 
52 & 5212 & +24 & 0911 & +180 &  4 & -53 &  4 & 26.9 & 0.6 & 19.41 & 0.03 & 17.85 & 0.01 & 15.59 & 0.01 & 11.38 & 0.02 & -1.26 & 14.5 & 0.144 & 8,9,10 \\ 
55 & 1910 & +17 & 5799 & +157 &  4 & -30 &  4 & 24.4 & 0.6 & 18.93 & 0.01 & 17.49 & 0.01 & 15.53 & 0.01 & 11.73 & 0.02 & -0.84 & 10.2 & 0.136 &  \\ 
55 & 2289 & +19 & 4965 & +163 &  5 & -32 &  5 & 25.2 & 0.8 & 20.15 & 0.04 & \multicolumn{2}{c}{---} & 16.40 & 0.01 & 12.24 & 0.02 & +0.20 & 11.0 & 0.108 &  \\ 
56 & 8816 & +12 & 0197 & +141 &  4 & -14 &  4 & 22.6 & 0.7 & 19.44 & 0.03 & 17.98 & 0.01 & 15.75 & 0.01 & 11.64 & 0.02 & -0.89 &  8.9 & 0.151 &  \\ 
57 & 7392 & +18 & 3019 & +183 &  5 & -32 &  5 & 29.7 & 0.8 & 20.71 & 0.03 & 19.04 & 0.01 & 16.64 & 0.01 & 11.78 & 0.02 & +0.76 & 14.2 & 0.112 & 8,1,9 \\ 
57 & 9864 & +41 & 2047 & +145 &  4 & -97 &  4 & 24.0 & 0.5 & \multicolumn{2}{c}{---} & 16.88 & 0.01 & 15.05 & 0.00 & 11.54 & 0.02 & -0.37 & 20.1 & 0.149 &  \\ 
58 & 5058 & +23 & 2761 & +192 &  5 & -66 &  5 & 32.2 & 0.8 & 21.00 & 0.04 & 19.49 & 0.01 & 16.83 & 0.01 & 12.00 & 0.02 & -0.80 & 16.7 & 0.098 & 11,12,13,14 \\ 
59 & 7924 & +9 & 2840 & +143 &  4 &  -1 &  4 & 24.3 & 0.7 & 18.41 & 0.01 & 17.12 & 0.01 & 15.19 & 0.01 & 11.59 & 0.03 & +0.31 &  9.4 & 0.145 &  \\ 
60 & 8216 & $-$1 & 8217 & +147 &  4 & +41 &  4 & 25.9 & 0.6 & 17.03 & 0.01 & 15.74 & 0.01 & 14.07 & 0.01 & 10.75 & 0.02 & +1.19 & 16.3 & 0.201 &  \\ 
62 & 7767 & +12 & 7967 & +125 &  5 & -18 &  5 & 23.0 & 0.9 & 19.24 & 0.02 & 17.88 & 0.01 & 15.76 & 0.01 & 11.84 & 0.03 & -0.53 &  5.4 & 0.137 & 15 \\ 
64 & 5574 & +25 & 6610 & +106 &  4 & -56 &  4 & 20.9 & 0.7 & 19.70 & 0.01 & 18.27 & 0.01 & 15.99 & 0.01 & 11.86 & 0.02 & -1.27 &  7.8 & 0.148 &  \\ 
64 & 6898 & +21 & 0318 & +108 &  4 & -53 &  4 & 21.7 & 0.7 & 19.71 & 0.02 & 18.36 & 0.01 & 16.12 & 0.01 & 12.21 & 0.02 & -2.97 &  3.9 & 0.124 & 10 \\ 
64 & 7741 & +17 & 5732 & +100 &  4 & -26 &  4 & 19.3 & 0.7 & 18.35 & 0.01 & 17.02 & 0.01 & 15.16 & 0.01 & 11.51 & 0.02 & +0.09 &  5.9 & 0.189 & 12 \\ 
65 & 2000 & +15 & 2358 & +161 &  5 & -59 &  5 & 32.4 & 0.9 & 14.44 & 0.01 & 13.36 & 0.01 & \multicolumn{2}{c}{---} &  8.63 & 0.02 & -3.76 & 15.6 & 0.446 & 16 \\ 
65 & 4349 & +20 & 4029 & +101 &  4 & -39 &  4 & 20.1 & 0.7 & 20.10 & 0.02 & 18.69 & 0.01 & 16.58 & 0.01 & 12.51 & 0.02 & -0.80 &  4.8 & 0.118 & 17,18 \\ 
65 & 4412 & +20 & 3957 & +95 &  5 & -33 &  5 & 18.6 & 1.0 & 21.21 & 0.06 & 19.79 & 0.02 & 17.30 & 0.01 & 12.93 & 0.03 & -0.07 &  8.1 & 0.107 & 10 \\ 
65 & 4566 & +19 & 4857 & +125 &  4 & -48 &  4 & 24.9 & 0.7 & 19.78 & 0.02 & 18.35 & 0.01 & 15.98 & 0.01 & 11.67 & 0.02 & -1.34 &  6.6 & 0.137 & 19,20,21,22 \\ 
66 & 5793 & +17 & 0506 & +97 &  4 & -37 &  4 & 20.3 & 0.8 & 18.94 & 0.01 & 17.59 & 0.01 & 15.60 & 0.01 & 11.91 & 0.03 & -2.30 &  3.0 & 0.149 & 17,18 \\ 
66 & 9046 & +20 & 9441 & +89 &  6 & -36 &  6 & 18.3 & 1.1 & 19.82 & 0.02 & 18.47 & 0.01 & 16.18 & 0.01 & 12.07 & 0.02 & -0.49 &  9.1 & 0.153 & 23,10 \\ 
67 & 8182 & +15 & 0034 & +105 &  4 & -20 &  4 & 21.9 & 0.9 & 18.69 & 0.01 & 17.33 & 0.01 & 15.39 & 0.01 & 11.71 & 0.02 & +0.77 &  1.6 & 0.151 & 12 \\ 
68 & 8065 & +20 & 1337 & +99 &  6 & -33 &  6 & 21.0 & 1.1 & 21.06 & 0.04 & 19.58 & 0.02 & 17.12 & 0.01 & 12.61 & 0.02 & +1.39 &  3.4 & 0.109 &  \\ 
68 & 9657 & +12 & 2556 & +93 &  6 & -20 &  6 & 20.5 & 1.2 & \multicolumn{2}{c}{---} & 19.11 & 0.04 & 16.81 & 0.01 & 12.68 & 0.03 & -1.43 &  4.6 & 0.109 &  \\ 
69 & 2422 & +23 & 2774 & +90 &  5 & -45 &  5 & 19.8 & 1.0 & 20.23 & 0.03 & 18.72 & 0.01 & \multicolumn{2}{c}{---} & 12.57 & 0.02 & +0.11 &  7.1 & 0.117 &  \\ 
69 & 7437 & +15 & 6614 & +91 &  6 & -26 &  6 & 20.4 & 1.3 & 19.44 & 0.02 & 18.12 & 0.01 & 16.00 & 0.01 & 12.20 & 0.03 & -0.40 &  3.5 & 0.131 & 24,17,18 \\ 
70 & 1052 & $-$4 & 6222 & +63 &  5 & +32 &  5 & 14.8 & 1.1 & 20.04 & 0.02 & 18.61 & 0.02 & 16.59 & 0.01 & 12.74 & 0.03 & +2.10 & 30.0 & 0.140 &  \\ 
71 & 0616 & +5 & 7326 & +77 &  5 &  +1 &  5 & 17.9 & 1.2 & 21.97 & 0.19 & 20.07 & 0.03 & 17.46 & 0.01 & 12.52 & 0.03 & -0.97 & 14.0 & 0.129 & 25,26 \\ 
71 & 5113 & +19 & 4683 & +76 &  6 & -41 &  6 & 18.8 & 1.2 & 19.52 & 0.02 & 18.22 & 0.01 & 16.09 & 0.01 & 12.30 & 0.02 & -2.08 &  8.2 & 0.135 &  \\ 
71 & 5402 & +18 & 9578 & +79 &  6 & -38 &  6 & 19.3 & 1.3 & 21.01 & 0.03 & 19.41 & 0.01 & 17.00 & 0.01 & 12.72 & 0.03 & -1.14 &  6.9 & 0.112 &  \\ 
71 & 6875 & +24 & 6112 & +90 &  6 & -61 &  6 & 22.1 & 1.2 & 19.73 & 0.02 & 18.43 & 0.01 & 16.14 & 0.01 & 12.13 & 0.02 & -1.12 &  7.2 & 0.126 & 10 \\ 
71 & 9778 & $-$0 & 7934 & +76 &  5 & +34 &  5 & 19.0 & 1.2 & \multicolumn{2}{c}{---} & 20.39 & 0.06 & 17.68 & 0.01 & 12.99 & 0.03 & +2.80 & 16.7 & 0.103 &  \\ 
72 & 1916 & +11 & 3857 & +97 &  4 & -25 &  4 & 24.0 & 1.0 & 19.11 & 0.01 & 17.73 & 0.01 & 15.57 & 0.01 & 11.53 & 0.03 & -2.02 &  7.3 & 0.150 &  \\ 
72 & 4634 & $-$16 & 5419 & +90 &  4 & +74 &  4 & 21.7 & 0.8 & \multicolumn{2}{c}{---} & \multicolumn{2}{c}{---} & 13.92 & 0.01 & 10.77 & 0.02 & -0.52 & 26.7 & 0.243 &  \\ 
73 & 4205 & $-$4 & 4760 & +82 &  4 & +40 &  4 & 21.1 & 1.0 & 17.66 & 0.01 & 16.33 & 0.01 & 14.70 & 0.01 & 11.35 & 0.02 & +0.69 & 18.0 & 0.187 &  \\ 
75 & 1623 & +24 & 3828 & +66 &  4 & -52 &  4 & 18.6 & 1.0 & 19.11 & 0.01 & 17.72 & 0.01 & 15.82 & 0.01 & 12.18 & 0.02 & -1.14 & 11.9 & 0.144 & 10 \\ 
75 & 1655 & +9 & 1261 & +52 &  6 &  -5 &  6 & 14.2 & 1.5 & 19.14 & 0.01 & 17.79 & 0.01 & 15.85 & 0.00 & 12.19 & 0.03 & -0.30 & 26.6 & 0.188 &  \\ 
75 & 8281 & +16 & 5163 & +58 &  6 & -26 &  6 & 16.6 & 1.5 & 19.00 & 0.01 & 17.70 & 0.01 & 15.70 & 0.01 & 12.00 & 0.02 & -0.60 & 15.7 & 0.174 &  \\ 
77 & 1714 & +14 & 7444 & +47 &  4 & -19 &  4 & 14.1 & 1.1 & 18.60 & 0.01 & \multicolumn{2}{c}{---} & 15.53 & 0.01 & 12.11 & 0.02 & -0.66 & 26.3 & 0.197 &  \\ 
77 & 9975 & +11 & 3283 & +56 &  5 & -18 &  5 & 17.7 & 1.6 & \multicolumn{2}{c}{---} & 19.09 & 0.02 & 16.75 & 0.01 & 12.21 & 0.02 & -1.76 & 14.7 & 0.149 &  \\ 
79 & 7312 & +8 & 4541 & +51 &  5 & -10 &  5 & 17.4 & 1.8 & 19.01 & 0.02 & \multicolumn{2}{c}{---} & 15.81 & 0.00 & 12.24 & 0.02 & -1.52 & 17.2 & 0.149 &  \\ 
79 & 8109 & +24 & 1261 & +47 &  5 & -43 &  5 & 16.1 & 1.3 & 18.82 & 0.01 & 17.49 & 0.01 & 15.63 & 0.01 & 11.93 & 0.02 & -0.03 & 20.5 & 0.187 &  \\ 
80 & 2844 & +19 & 6154 & +48 &  5 & -23 &  5 & 15.0 & 1.5 & 19.00 & 0.01 & 17.57 & 0.01 & 15.76 & 0.00 & 12.25 & 0.02 & +2.67 & 24.0 & 0.172 &  \\ 
80 & 4052 & +1 & 7577 & +53 &  4 &  +8 &  4 & 17.9 & 1.3 & \multicolumn{2}{c}{---} & 16.48 & 0.01 & 14.74 & 0.01 & 11.28 & 0.03 & -1.71 & 20.0 & 0.233 &  \\ 
80 & 4142 & +1 & 7491 & +51 &  4 &  +6 &  4 & 17.1 & 1.3 & \multicolumn{2}{c}{---} & 14.76 & 0.01 & 13.35 & 0.01 & 10.27 & 0.02 & -2.19 & 21.6 & 0.403 &  \\ 
80 & 5064 & +15 & 5819 & +45 &  5 & -22 &  5 & 15.8 & 1.7 & 19.14 & 0.01 & 17.78 & 0.01 & 15.86 & 0.01 & 12.24 & 0.03 & -0.12 & 20.9 & 0.165 &  \\ 
80 & 6989 & +43 & 6304 & +63 &  4 & -108 &  4 & 20.4 & 0.6 & 16.29 & 0.01 & 15.11 & 0.01 & 13.59 & 0.01 & 10.41 & 0.02 & +1.69 & 24.3 & 0.314 &  \\ 
80 & 9093 & $-$14 & 0506 & +107 &  6 & +147 &  6 & 42.3 & 1.3 & \multicolumn{2}{c}{---} & \multicolumn{2}{c}{---} & 17.60 & 0.01 & 11.64 & 0.03 & +1.48 & 29.7 & 0.091 & 13,27,28,29 \\ 
81 & 2259 & +42 & 1577 & +37 &  4 & -105 &  4 & 18.4 & 0.7 & 16.50 & 0.01 & 15.23 & 0.01 & 13.62 & 0.01 & 10.37 & 0.02 & -3.86 & 25.7 & 0.357 &  \\ 
82 & 7044 & +1 & 9849 & +38 &  4 & +11 &  4 & 15.2 & 1.5 & \multicolumn{2}{c}{---} & \multicolumn{2}{c}{---} & 15.37 & 0.01 & 12.05 & 0.03 & -0.12 & 28.2 & 0.187 &  \\ 
83 & 3414 & +34 & 2936 & +31 &  4 & -84 &  4 & 18.1 & 0.9 & 18.54 & 0.01 & 17.30 & 0.01 & \multicolumn{2}{c}{---} & 11.86 & 0.02 & -3.74 & 21.9 & 0.171 &  \\ 
84 & 2374 & +8 & 7568 & +33 &  6 & -10 &  6 & 15.3 & 2.4 & 19.81 & 0.02 & 18.37 & 0.01 & 16.46 & 0.00 & 12.67 & 0.03 & -1.55 & 26.1 & 0.140 &  \\ 
89 & 1518 & +12 & 1363 & +20 &  4 & -15 &  4 & 14.9 & 2.2 & 17.82 & 0.01 & 16.52 & 0.01 & 14.97 & 0.01 & 11.73 & 0.02 & -0.56 & 29.7 & 0.226 &  \\ 
90 & 7672 & +36 & 1077 & +42 &  5 & -120 &  5 & 26.2 & 1.0 & 16.09 & 0.01 & 14.81 & 0.01 & 13.13 & 0.01 &  9.84 & 0.02 & +2.10 & 22.3 & 0.320 &  \\ 
96 & 1005 & +40 & 5209 & +21 &  5 & -120 &  5 & 22.6 & 0.9 & 15.08 & 0.01 & 13.96 & 0.01 & \multicolumn{2}{c}{---} &  9.52 & 0.02 & +3.01 & 27.0 & 0.428 &  \\ 
98 & 5462 & +35 & 8668 &  -6 &  4 & -95 &  4 & 20.3 & 0.9 & 17.90 & 0.01 & \multicolumn{2}{c}{---} & \multicolumn{2}{c}{---} & 11.42 & 0.02 & -0.80 & 28.0 & 0.188 &  \\ 
98 & 7939 & +29 & 2054 &  -2 &  4 & -68 &  4 & 18.4 & 1.1 & 16.26 & 0.01 & \multicolumn{2}{c}{---} & 13.64 & 0.01 & 10.50 & 0.02 & +0.13 & 28.5 & 0.337 &  \\ 
102 & 4987 & +39 & 4349 & -24 &  5 & -120 &  5 & 23.4 & 0.9 & 15.60 & 0.01 & 14.39 & 0.01 & \multicolumn{2}{c}{---} &  9.84 & 0.02 & -1.11 & 29.4 & 0.358 &  \\ 

\label{NewCand}
\end{longtable}
\tablebib{
(1)~\citet{Cruz02}; (2)~\citet{PhanB06}; (3)~\citet{West08}; (4)~\citet{Galvez10}; (5)~\citet{Tinne93}; (6)~\citet{Bocha05}; (7)~\citet{Deaco09}; (8)~\citet{Gizis00}; (9)~\citet{Reid02h}; (10)~\citet{Slesn06}; (11)~\citet{Kirkp97}; (12)~\citet{Bouvi08}; (13)~\citet{Cruz03}; (14)~\citet{Close03}; (15)~\citet{Cruz07}; (16)~\citet{Gicla62}; (17)~\citet{Reid92}; (18)~\citet{Reid00h}; (19)~\citet{Bryja94}; (20)~\citet{Siegl03}; (21)~\citet{Watso09}; (22)~\citet{Dupuy11}; (23)~\citet{Luhma06sd}; (24)~\citet{Bryja92}; (25)~\citet{Reid08X}; (26)~\citet{Faher09}; (27)~\citet{Reid08}; (28)~\citet{Seifa10}; (29)~\citet{Blake10}.
}
}
\end{landscape}
\end{center}

\end{document}